\documentclass[prb, reprint, amsmath,amssymb,aps,floatfix,superscriptaddress]{revtex4-2}

\usepackage{graphicx}
\usepackage{hyperref}
\usepackage{color}
\usepackage{bm}
\usepackage{amssymb}
\usepackage{amsmath}
\usepackage[normalem]{ulem}
\usepackage[T1]{fontenc}
\usepackage{multirow}
\usepackage{xcolor}
\usepackage{chemformula}

\newcommand{\lc}{\lowercase}
\begin{document}

\title{Non--Fermi liquid behavior of doped Kondo insulator: The unique properties of CeRhSb$_{1-x}$Te$_x$. }

\author{A.~\'{S}lebarski}
\affiliation{Institute of Low Temperature and Structure Research, Polish Academy of Sciences, ul. Ok\'{o}lna 2, 50-422 Wroc{\l}aw, Poland}
\affiliation{Centre for Advanced Materials and Smart Structures, 
Polish Academy of Sciences, ul. Ok\'{o}lna 2, 50-422 Wroc\l aw, Poland}
\affiliation{$^{\star}$ Author to whom correspondence should be addressed: andrzej.slebarski@us.edu.pl}
\author{J. Spa\l ek}
\affiliation{Institute of Theoretical Physics, Jagiellonian University, ul. \L{}ojasiewicza 11, 30-348 Krak\'ow, Poland}
\author{M. Fija\l kowski}
\affiliation{Institute of Physics,
University of Silesia in Katowice, ul. 75 Pu\l ku Piechoty 1, 41-500 Chorz\'ow, Poland}
\affiliation{Centre for Advanced Materials and Smart Structures, 
Polish Academy of Sciences, ul. Ok\'{o}lna 2, 50-422 Wroc\l aw, Poland}
\begin{abstract}

It follows from our analysis of CeRhSb that the formation of Kondo insulator state due to the presence of the collective spin singlet state is strongly reduced by its doping with various dopants when their amount exceeds 8--10\%, regardless of whether they are substituted for Ce, Rh or Sb. 
A wide variety of experimental results (electrical resistivity $\rho$, magnetic susceptibility $\chi$, specific heat $C$, x-ray photoelectron spectroscopy) and theoretical investigations have convincingly demonstrated the proposed earlier scaling law $\chi\times\rho=const.$ in the Kondo insulator regime, which is universal for all known Kondo insulators. We also analyze the properties of the Griffiths--phase for CeRhSb when Pd substitutes Rh, or  Sb is fractionally replaced by Te and Sn, whereas doping of Ce with La leads to the formation of magnetic cluster structure as a result of the Kondo hole effect.  Magnetoresistance of CeRhSb and CeRhSb$_{0.98}$Te$_{0.02}$ as a function of the field $B$  shows a $-B^2$ behavior, which provides evidence for the topologically nontrivial nature of these compounds, as was previously  predicted theoretically for CeRhSb on the basis on the band structure calculations. 

\end{abstract}
\maketitle

\newpage

\section{Introduction}

The stoichiometric CeRhSb  Kondo  insulator  (KI)  belongs  to  the  class  of  strongly  correlated  materials  with the small-gap $\Delta \sim 8$ K semiconducting properties
and  a  heavy-fermion  metallic  state,  setting  in gradually at  elevated  temperature  $T \geq  \Delta$ \cite{Aeppli1992}.  
Within the framework of the periodic Anderson model, the gap arises from the strongly renormalized by electronic correlations and hybridization $V$ between the conduction band and the strongly correlated $f$-electron states. Moreover,  $\Delta$ is strongly temperature dependent \cite{Ekino1995,Spalek2000}, in contrast to usual narrow-gap semiconductors.  
A full understanding  of  KIs  is  in part obscured by the circumstance that relatively small values of both magnetic susceptibility $\chi$  and the  gap  when  $T\rightarrow 0$  lead  to  their  properties alteration by  a  minute  content  of  magnetic  or  other  impurities \cite{Takabatake1998,Slebarski1998,Slebarski1996}.  In effect, it is often difficult to differentiate between the intrinsic and extrinsic properties of  these  systems.  It has been shown that even a relatively small number of in-gap states due to the lattice defects or doping plays a destructive role in the formation of the Kondo insulating ground state, manifested, e.g., in resistivity $\rho$ through its deviation from the activated behavior at the lowest temperatures.  
Here we show that the atomic disorder, resulting from lattice defects  in parent CeRhSb  or generated by doping,  can lead to local magnetic inhomogeneities, characteristic of the onset of the Griffiths phase. 

For the last two decades, an increasing attention has been devoted to Kondo insulators due to the fact that they can also become topological Kondo insulators (TKI), as they possess that extremely narrow energy gap. One of the first arguments was the observation of  deviation from activated behavior  of resistivity at the lowest temperatures $T\ll \Delta$. The first theoretically predicted TKI was SmB$_6$ \cite{Dzero2010,Dzero2012}. The metallic 
surface states have been documented by angle-resolved photoemission spectroscopy (ARPES)
\cite{Neupane2013,Xu2013,Jiang2013} and electrical transport measurements 
\cite{Wolgast2013,Kim2013,Kim2014}, providing an evidence for  SmB$_6$  as a possible TKI. It has also been predicted theoretically that CeNiSn \cite{Chang2017} and CeRhSb 
\cite{Nam2019} may be regarded as novel topological Kondo insulators possessing M\"obius-twisted surface states. These materials were classified as the Kondo insulators with the gap that closes along $a$ axis \cite{Nakamoto1995}. In effect, CeNiSn and CeRhSb have been
termed as {\it failed Kondo insulators} \cite{Chang2017} with a pseudogap. But, the topological Fermi-surface features have not been confirmed experimentally as yet \cite{Seong2019}.
We observe a negative magnetoresistance $MR$ with quadratic field dependence for CeRhSb which may indicate the chiral magnetic behavior, suggesting the topologically nontrivial electronic state of that material.

The structure of this paper is as follows. After Introduction and Experimental Details (Secs. I and II, respectively), in Sec. III, we specify, on the example of \ch{CeRh_{1-x}Pd_xSb}, the critical doping $x$, at which the collective Kondo spin--singlet state (KI regime) transforms into metallic (non--KI) state (cf. Fig. \ref{fig:DELTA_vs_x_CeRh-PdSb_GS}). For reference, we also specify there first the detailed resistivity behavior (cf. Fig \ref{fig:Fig_R_Te-02}), and magnetic  crossover from anomalous to Curie--Weiss regime (cf. Figs. \ref{fig:Fig_CeRhSb_CHI-R_skalowanie} and \ref{fig:Chi_ac_f_CeLa02RhSb}). This anomalous magnetic behavior is compared with the calculated temperature behavior of a canonical Kondo insulator (cf. also Appendix A for some theoretical details). On the basis of electrical resistivity $\rho(T)$ scaling, we have determined the anomalous applied field scaling of the pseudogap $\Delta$ for \ch{CeRhSb_{1-x}Te_x}, as well as the temperature scaling of the magnetic susceptibility $\chi$ and relative specific heat $C/T$ (cf. Figs. \ref{fig:Fig_Gap-vs-B_x}-\ref{fig:Fig_M-H_Te_02_fit}, respectively). The data extrema are associated with the onset of Griffits phase in the temperature range $T\sim 1$ K. All of them provide further evidence for non--KI behavior for \ch{CeRhSb_{1-x}Te_x}. The data presented in Sec. IIIC elaborate on the details of the proposed non--KI scenario. The discussed data in Sec. IIIA--C have provided novel scaling laws accumulated in Table \ref{tab:C_CHI}. Finally, Sec. IV contains a brief recapitulation of our analysis. 

\section{Experimental details}
 
\begin{figure}[ht]
\centering
\includegraphics[width=0.43\textwidth]{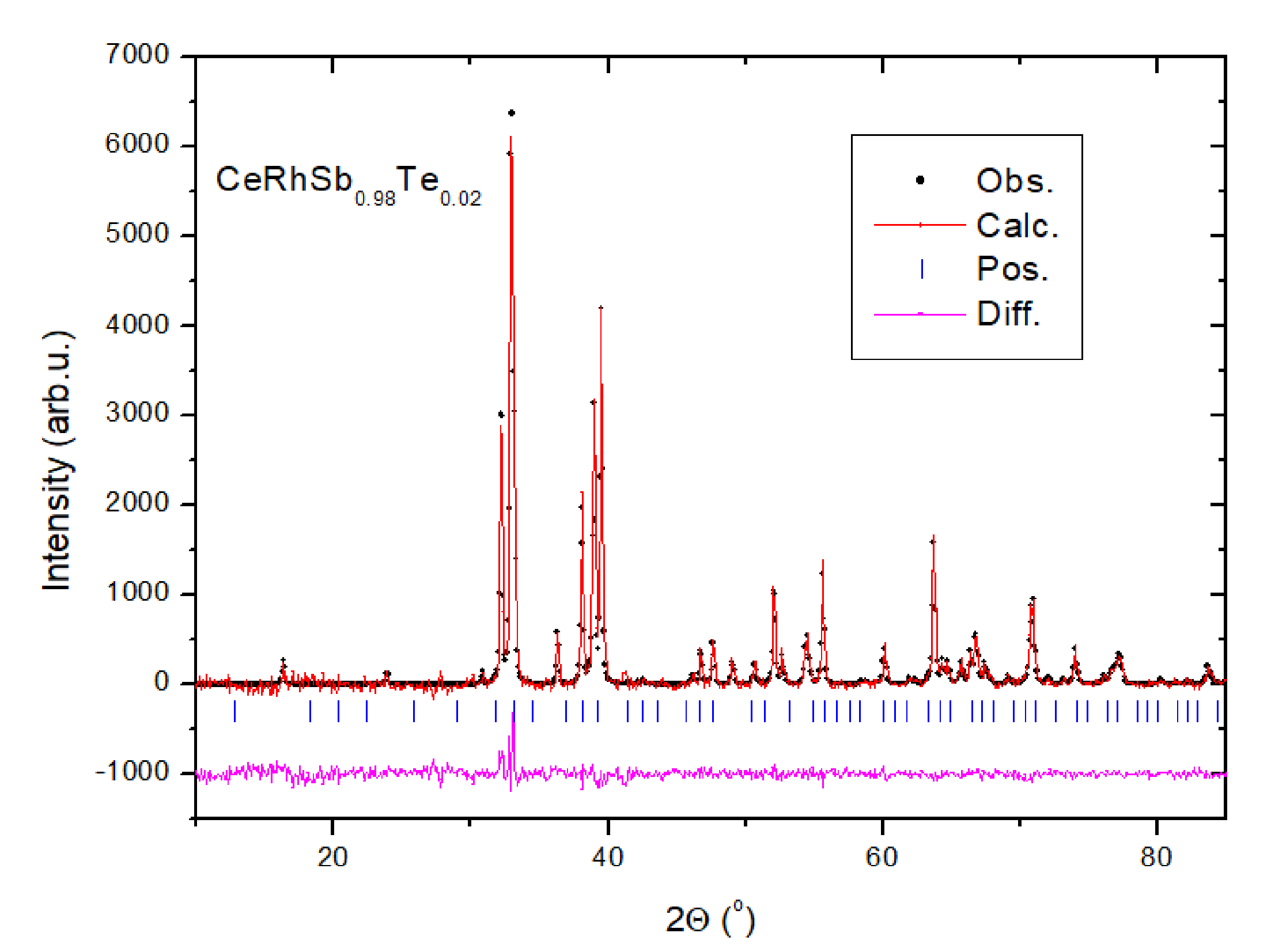}
\caption{\label{fig:XRD}
Powder XRD pattern of  CeRhSb$_{0.98}$Te$_{0.02}$ recorder at room temperature. The solid red line (Calc.) through the experimental points (black points - Obs.) is a calculated Rietveld refinement profile. The lowest curve (Diff.) represents the difference between the experimental and calculated results. The vertical bars (Pos) indicate the Bragg positions for the $\epsilon$-TiNiSi-type  structural model. The weighted profile $R_{wp}$ factor is 3.2\%,  $R_{exp}= 3.0$\%, and the lattice parameter $a=7.4265(2)$ \AA~(7.4130 \AA), $b=4.6199(0)$ \AA~(4.6124 \AA), and $c=7.8604(3)$ \AA~(7.8531 \AA) are compared with the lattice parameters of CeRhSb (in the brackets). The meaning  of  $R$  is  defined, e.g., in Ref. \cite{Toby2006}.} 
\end{figure}

Polycrystalline samples of Ce$_{1-x}$La$_x$RhSb, CeRh$_{1-x}$Pd$_x$Sb, CeRhSb$_{1-x}$Sn$_x$, and CeRhSb$_{1-x}$Te$_x$ were prepared via arc melting. 
For the first three series mentioned, pure CeRhSb and  respectively LaRhSb ($\epsilon$-TiNiSi-type structure, space group $Pnma$), CePdSb (CaIn$_2$-type structure, space group P$6_3/mmc$ \cite{Malik1991,Slebarski2006a}), or CeRhSn (Fe$_2$P-type structure, space group P$\bar{6}2m$ \cite{Slebarski2002a}) 
were the first arc melted, and then the respective  diluted alloys  were prepared by mixing the nominal compositions of the master compounds.
CeRhTe is, however, not formed, therefore the stoichiometric amounts of the constituent elements  were arc melted for CeRhSb$_{1-x}$Te$_x$.
To ensure homogeneity, each sample was turned over and remelted several times and then annealed at 800 $^o$C for 2 weeks. An excess amount of 1\% Sb was added to compensate for losses of this  metal within the melting process. 
The crystal structure of the annealed samples was checked by powder x-ray diffraction. The Rietveld  structural refinement revealed that the samples crystallize with  orthorhombic the $\epsilon$-TiNiSi structure (space group $Pnma$) and are obtained as single phases in a limited concentration $x<\sim 0.1$ range. The series of   CeRhSb$_{1-x}$Te$_x$ samples have not been known and investigated so far. The single--phase components of the series were obtained for $x<0.08$. The x-ray diffraction pattern of CeRhSb$_{0.98}$Te$_{0.02}$ is shown in Fig. \ref{fig:XRD}.
 
The ac electrical resistivity $\rho$ and ac magnetic susceptibility $\chi_{ac}$  measurements were performed using the quantum design (QD) PPMS platform. 
The resistivity and specific heat was measured in the temperature range $0.4-300$ K and in external magnetic fields up to 9 T also using QD PPMS platform. dc magnetic susceptibility and the  magnetization measurements were carried out using a commercial QD SQUID magnetometer in the magnetic field up to 7 T.

\section{Interplay between spin-glass-like and  Kondo insulating state: The case of local  disorder in C\lc{e}R\lc{h}S\lc{b}}
\subsection{Universal scaling law: $\chi (T)\rho (T)= const.$}  

\begin{figure}[ht]
\includegraphics[width=0.48\textwidth]{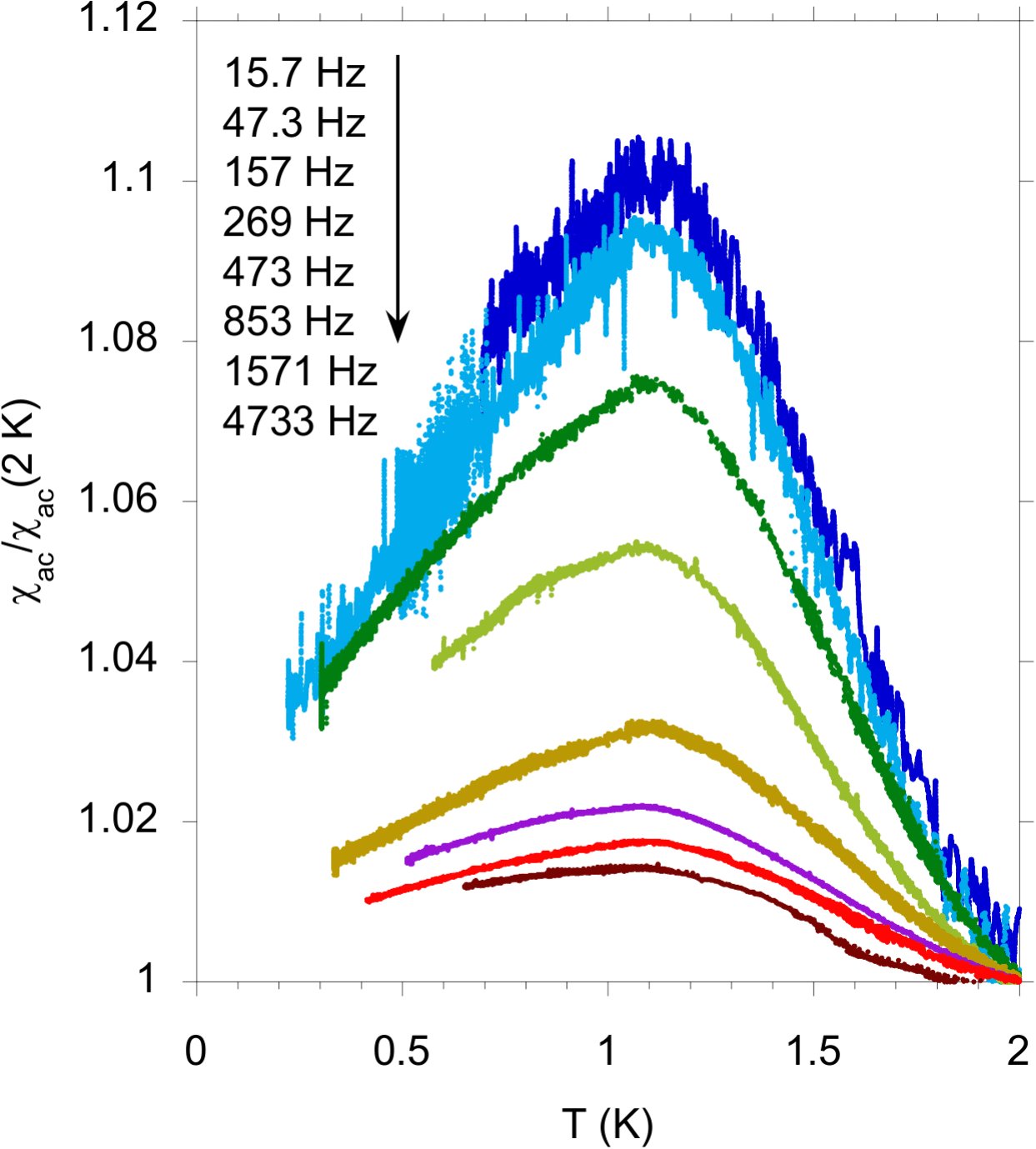}
\caption{\label{fig:Chi_ac_f_CeLa02RhSb}
Low-temperature ac magnetic susceptibility for Ce$_{0.98}$La$_{0.02}$RhSb, measured at different frequencies $\nu$ under magnetic field of 2 G. The data are normalized $\chi_{ac}$ at 2 K, respectively (taken from Ref. \cite{Slebarski2010}). 
} 
\end{figure}

Our  earlier  research on CeRhSb \cite{Slebarski2010}, CeRhSb$_{1-x}$Sn$_x$ \cite{Slebarski2005,Spalek2005}, CeNiSn \cite{Slebarski2008},  and  CeNi$_{1-\delta}$Sn$_{1+\delta-x}$Sb$_x$ \cite{Slebarski2009a}, as  well  as  that  on  Ce$_3$Bi$_4$Pt$_3$  \cite{Spalek2011,Hundley1990}  and  FeSi  \cite{Spalek2011,Jaccarino1967} follow that the Kondo insulating state can be characterized by the following features:  (i) The  number  of  valence  electrons  is even, i.e.,  it  is  18  for CeRhSb and CeNiSn (including  one  4f  electron  due  to  Ce),  whereas  Ce$_3$Bi$_4$Pt$_3$  has  54  such electrons and FeSi has 12. Based on that fact one can expect the complete screening of the atomic moment of either Ce$^{3+}$  or  Fe$^{4+}$ ions within  the formula  unit  if  the coupling  of  the  surrounding  screening  carriers  has an  antiferromagnetic  character. (ii) The  resistivity $\rho (T)$  is    thermally  activated  in  the  limited  temperature  range  with  a  small  activation gap,  $\Delta\sim(1-10)$ meV. However, the gap  is sometimes not easy to detect as the {\it shallow} impurity in-gap states form an impurity band, overlapping  with the upper and/or the lower hybridized band, thus leading to a semimetallic behavior. (iii) The narrow  gap or pseudogap is associated with the collective Kondo spin-singlet state formation, singled out directly by the decreasing magnetic susceptibility $\chi (T)$ when $T \rightarrow 0$, provided such a system does not order magnetically. In that case, the impurity contribution must be singled out and properly subtracted. Also, the disorder creates localized states in the pseudogap region and leads to a magnetic glassy state, which gives rise to the  Curie-Weiss contribution $y\frac{C}{T-\theta_{CW}}$  to $\chi(T)$ at the lowest temperatures with $y\sim 0.01$, as well as removes the activated behavior when the dopant content  is $x>\sim 0.08$. (iv) The  resistivity  and  magnetic  susceptibility  obey  \cite{Slebarski2005} a  simple  scaling  law  $\chi(T)\rho(T)=const.$, 
particularly in the regime $k_BT < \Delta$, when the number of excited carriers is small. The scaling law $\chi\rho = const.$ is universal  for the known Kondo insulators, provided that impurity contribution is properly subtracted, see e.g., Refs. \cite{Slebarski2005,Slebarski2008,Spalek2011}. It is worth noting that for CeRhSb of very high purity \cite{Takabatake1995} the susceptibility rapidly decreases  in the  KI regime and exhibits a clear $\chi\rightarrow 0$ behavior for $T\rightarrow 0$, cf., Ref. \cite{Slebarski2005}.
\begin{figure}[ht]
\includegraphics[width=0.4\textwidth]{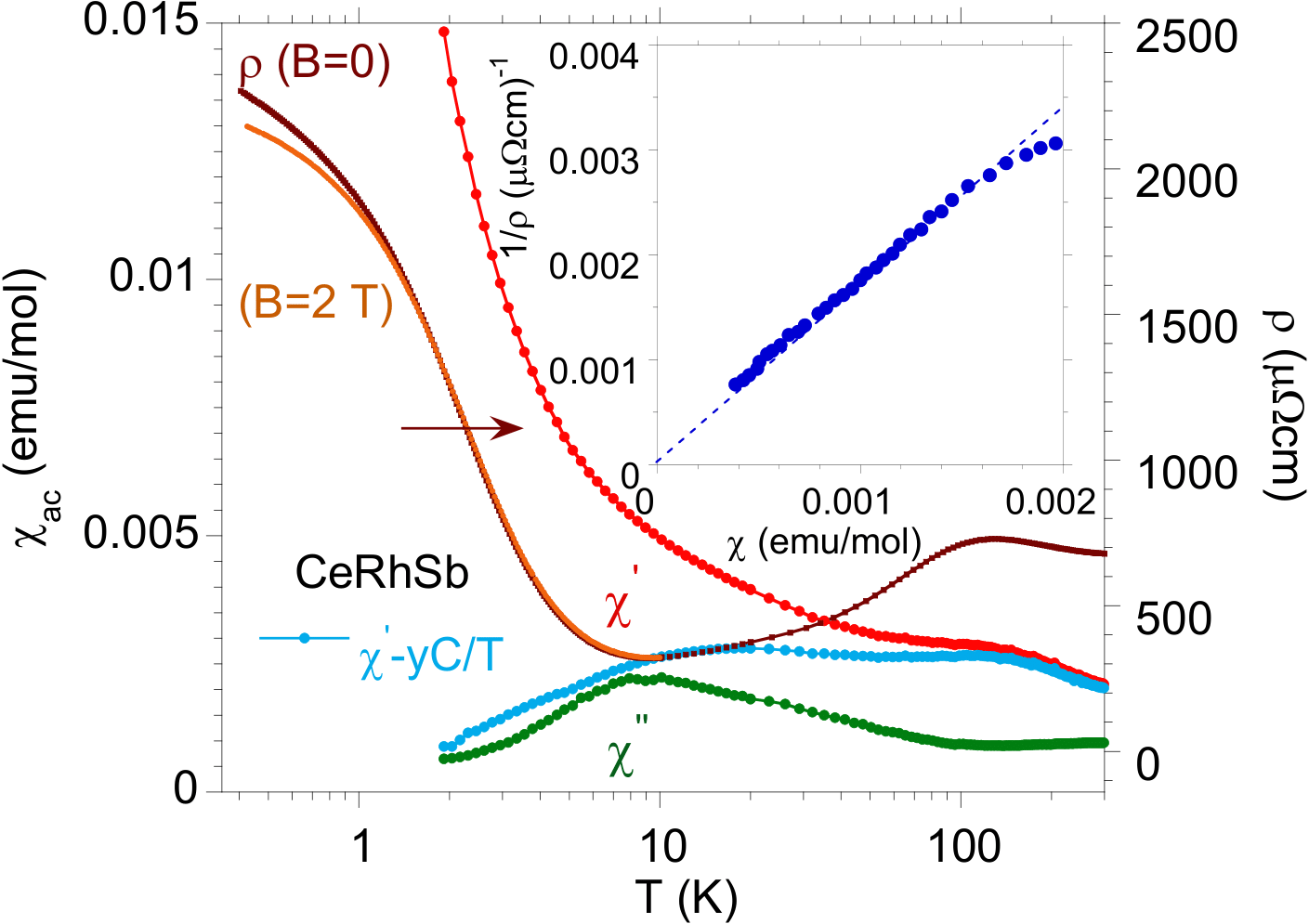}
\caption{\label{fig:Fig_CeRhSb_CHI-R_skalowanie}
Temperature dependence of the real ($\chi'$, red points) and imaginary ($\chi''$, green points) components of ac magnetic susceptibility for CeRhSb ($\nu=1$ kHz, amplitude of the magnetic field $B=2$ G). The blue points represent $\chi'(T)$ subtracted by the Curie-Weiss contribution $y\frac{C}{T-\theta_{CW}}$, where $y=0.028$ and $\theta_{CW}=0.3$ K are from the best fit of C-W expression to $\chi'$ for $T<100$ K. The brown and yellow points show the resistivity for CeRhSb at $B=0$ and 2 T, respectively. The inset demonstrates the linear scaling law between $\chi'-yC/T$ and $1/\rho$ in the temperature range 
 ($1.8 <T<5$) K. 
} 
\end{figure}
The  scaling $\chi (T)\rho (T)=const.$ is the most characteristic behavior. Therefore, we discuss first the full-gap  semiconductor CeRhSb with  the  conductivity  gap  $\Delta  =  6$  K, obtained from fitting of the resistivity data to $\rho\sim e^{\Delta/k_BT}$ law. The  scaling  law $\chi (T)\rho (T)=const.$ shown in the inset to Fig.  \ref{fig:Fig_CeRhSb_CHI-R_skalowanie} is  clearly  obeyed. Additionally, the displayed there $\chi''$ component  exhibits a  broad 
maximum at $\sim 9$ K, which indicates  both the presence of the hybridization gap and of a very weak inhomogeneous  spin-glass-like  ground state.    
When the number of local defects increases, e.g., due to doping, then  $\chi_{ac}$   is  both frequency ($\nu$) and field dependent \cite{Slebarski2010}, as shown in Fig. \ref{fig:Chi_ac_f_CeLa02RhSb}  for the Kondo-hole Ce$_{0.98}$La$_{0.02}$RhSb semimetal. 
In the  case of the increased number of structural defects, the  scaling $\chi \rho = const.$ is practically impossible to be identified and is  due to the presence of a glassy phase, which  represents a strong disturbance of the paramagnetic KI ground state in this case.

\begin{figure}[ht]
\includegraphics[width=0.48\textwidth]{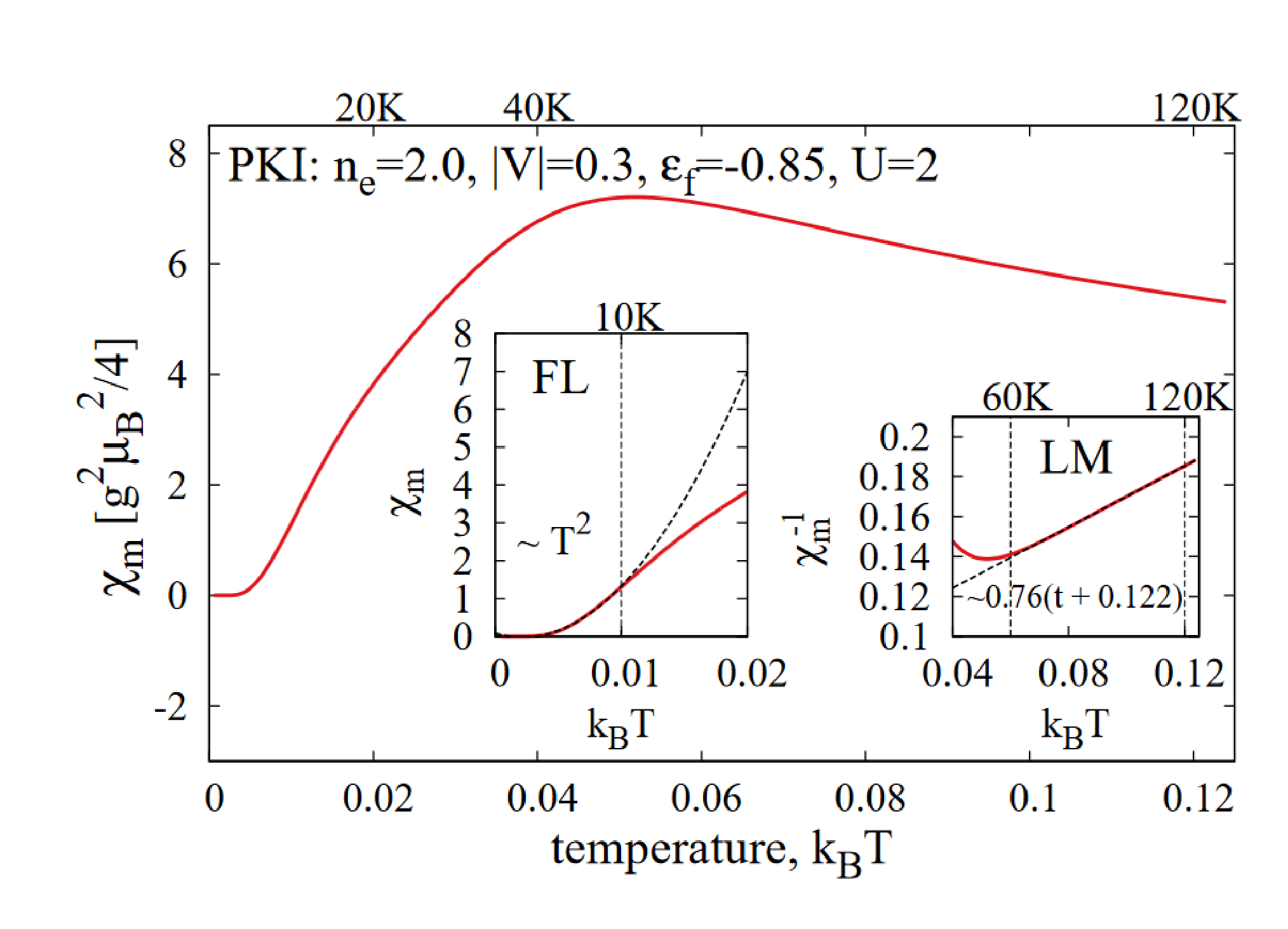}
\caption{\label{fig:PKI}
Temperature dependence of the zero-field magnetic susceptibility for the nonmagnetic  Kondo insulator (PKI). The insets show PKI
evolution from the Fermi-liquid regime (FL), to the localized moment regime (LM).
The upper-temperature scale is provided for bare conduction bandwidth $W = 103$ K (taken from Refs. \cite{PhD-Howczak, HowczakSpalek}; for details see Appendix A). 
} 
\end{figure}

\subsection{Nonmagnetic Kondo-compensated
state in the case with intraatomic hybridization: The $\chi\rightarrow 0$  for $T\rightarrow 0$ behavior within the periodic Anderson-Kondo model}  

To set a reference point for further discussion we have provided in Fig. \ref{fig:PKI} temperature dependence of the static magnetic susceptibility obtained theoretically for nonmagnetic KI \cite{PhD-Howczak}. The calculation involved an intraatomic hybridization $V$, as well as have been carried out for a number of electrons $n_e=2$ per site, the position of the $f$-level $\epsilon_f=-0.85$, and the $f-f$ Hubbard interaction magnitude $U=2$ (all parameters are in units of bare conduction-band bandwidth $W$). We see that for the even number of electrons the susceptibility $\chi_m\rightarrow 0$ with $T\rightarrow 0$, with $T^2$ dependence as for Landau Fermi liquid, whereas it takes the standard Curie-Weiss form in the high-temperature limit (cf. insets). Also, a broad maximum is observed around temperature $V^2/U\sim 0.05 W$, characteristic of a gradual transition from the Fermi-liquid (FL) to the $f$-electron localized moment (LM) state (with the $f$-level occupancy $n_f\rightarrow 1$ in the latter limit).

All these features, together with the universal scaling law
$\chi_m\times\rho = const.$, proposed and experimentally documented  earlier \cite{Slebarski2005,Spalek2005,Spalek2011},
provide us with the confidence that our method of approach describes properly the
the universal trend of the experimental data for KI systems.

\subsection{Non-Fermi-liquid behavior in CeRhSb with dopants: Griffits phase in CeRhSb$_{1-x}$Te$_x$, CeRh$_{1-x}$Pd$_x$Sb and Ce$_{0.98}$Th$_{0.02}$Sb}

\begin{figure}[ht]
\includegraphics[width=0.48\textwidth]{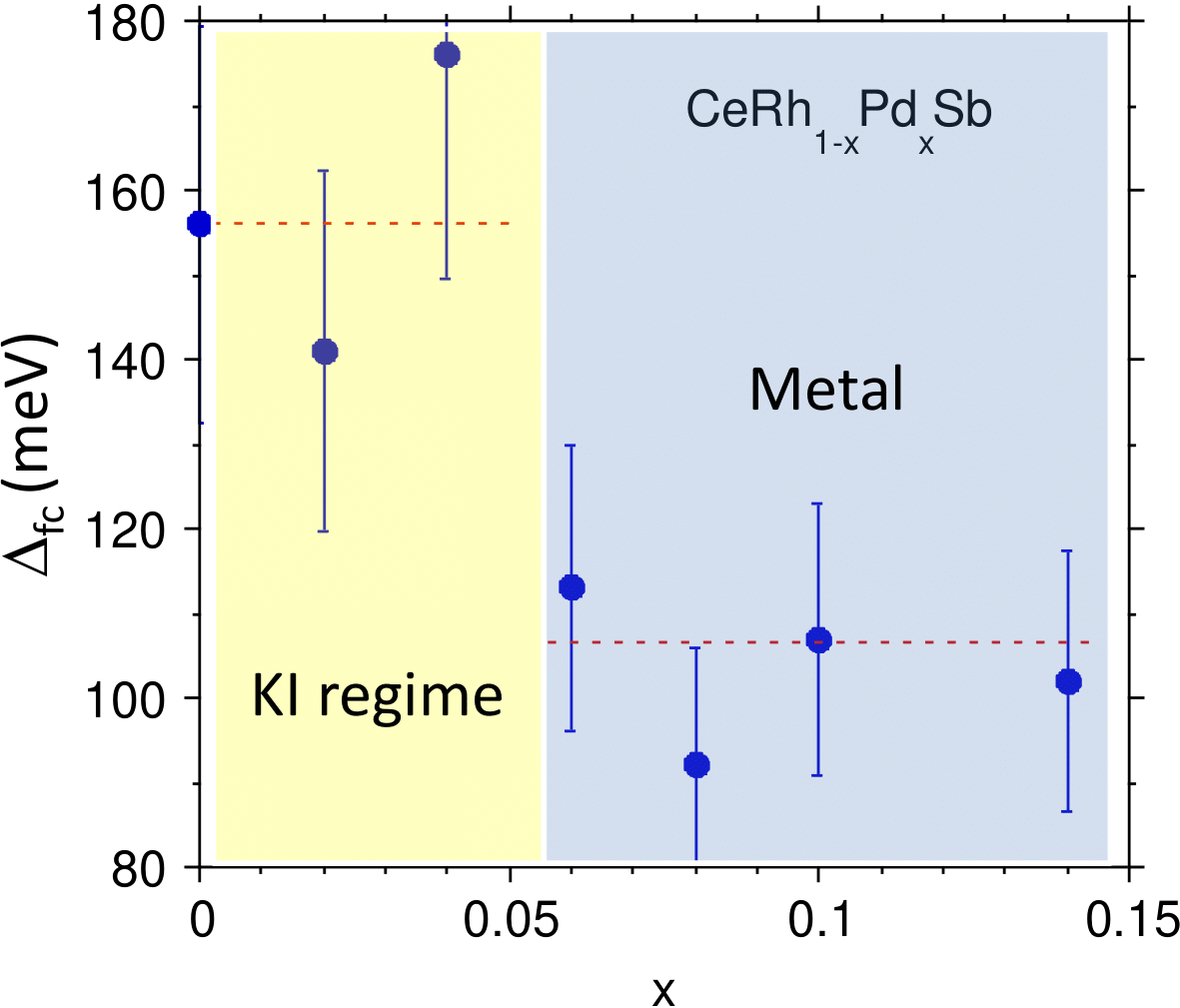}
\caption{\label{fig:DELTA_vs_x_CeRh-PdSb_GS}
CeRh$_{1-x}$Pd$_x$Sb: hybridization energy $\Delta_{fc}
=\pi V^2N(\epsilon_F)$ \cite{Anderson1961} vs $x$ obtained from the Ce $3d$ XPX spectra based on the Gunnarsson and Sch\"{o}nhammer theoretical model \cite{Fuggle1983},  $V$ is a matrix element of mixing
between $4f$ and conduction electrons at the Fermi level, $\epsilon_F$, and  $N(\epsilon_F)$ is the density of states at $\epsilon_F$. 
} 
\end{figure}

\begin{figure}[ht]
\includegraphics[width=0.48\textwidth]{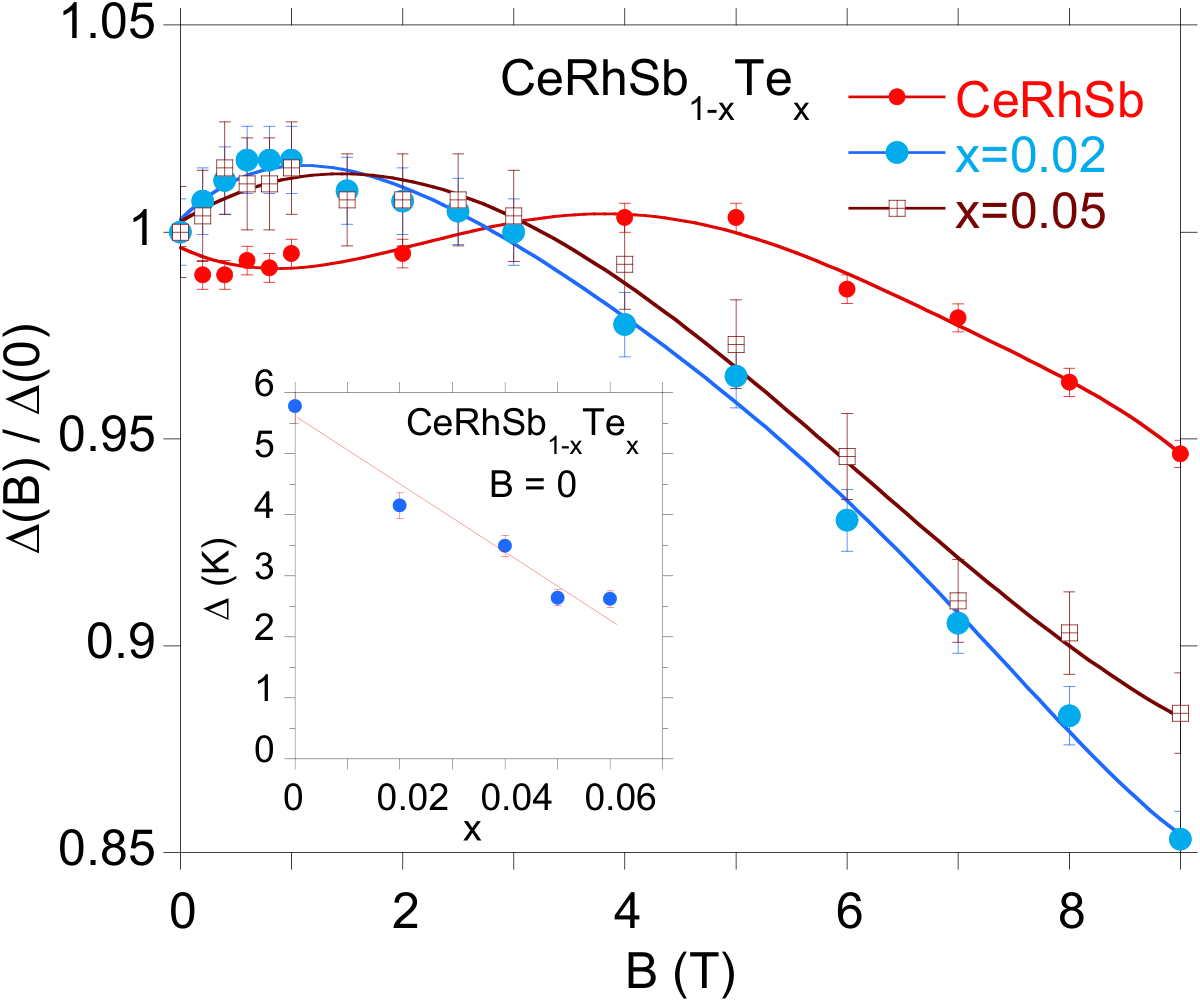}
\caption{\label{fig:Fig_Gap-vs-B_x}
CeRhSb$_{0.98}$Te$_{0.02}$; the KI gap $\Delta(B)$ as a function of applied magnetic field for components $x=0.02$ and 0.05 in comparison to the change of $\Delta(B)$ for parent compound CeRhSb. The inset displays the change of $\Delta$ with $x$ under zero magnetic field.
} \end{figure}

Doping of CeRhSb, regardless of whether the Ce, Rh or Sb atoms are replaced with another  atom $M$, causes the parent system ceases to be a single phase already for the impurity concentration $x$ larger than about 15--20\%. For the systems investigated  so far Ce$_{1-x}M_x$RhSb ($M = $La \cite{Malik1995,Adroja1996,Slebarski2010} or  Th \cite{Kim2003}), CeRh$_{1-x}M_x$Sb ($M = $Pd \cite{Menon1998,Slebarski2010a}), or CeRhSb$_{1-x}$Sn$_x$ \cite{Slebarski2005,Spalek2005,Slebarski2010,Slebarski2002a}, there is a critical concentration of $x_{cr}$ for each of them, which separates the Kondo-insulator phase from the metallic one. The reason for the  transformation KI$\rightarrow$metal with  increasing $x$  is  formation of partially filled hybridized band above critical concentration $x_{cr}$ as shown in Fig. \ref{fig:DELTA_vs_x_CeRh-PdSb_GS} for  the  CeRh$_{1-x}$Pd$_x$Sb alloys (for details, see Ref. \cite{Slebarski2010a}). 
This  transformation was predicted theoretically \cite{Doradzinski1997} and has been documented experimentally \cite{Slebarski2010,Slebarski2010a}. The effect is not dependent on whether the doping generates a Kondo hole (La, Th dopants)  and is neither associated with an increasing   (Pd, Te, dopants) nor with decreasing (Sn dopant) a number of carriers. Here  we  analyze  the impact of  Te for Sb substitution on the hybridized structure formation. Namely, we document  experimentally that the Te impurities (when $x<0.08$) generate the Griffits-type  phase as a result of doping--induced lattice disorder. Intriguingly, we  observe a similar Griffits-type phase  in CeRhSb doped by various other $M$ dopants. Below we present some of those examples.
First, we present the magnetic and electrical transport  data for CeRhSb with Sb substituted by Te, i.e., for CeRhSb$_{0.98}$Te$_{0.02}$.
\begin{figure}[ht]
\includegraphics[width=0.48\textwidth]{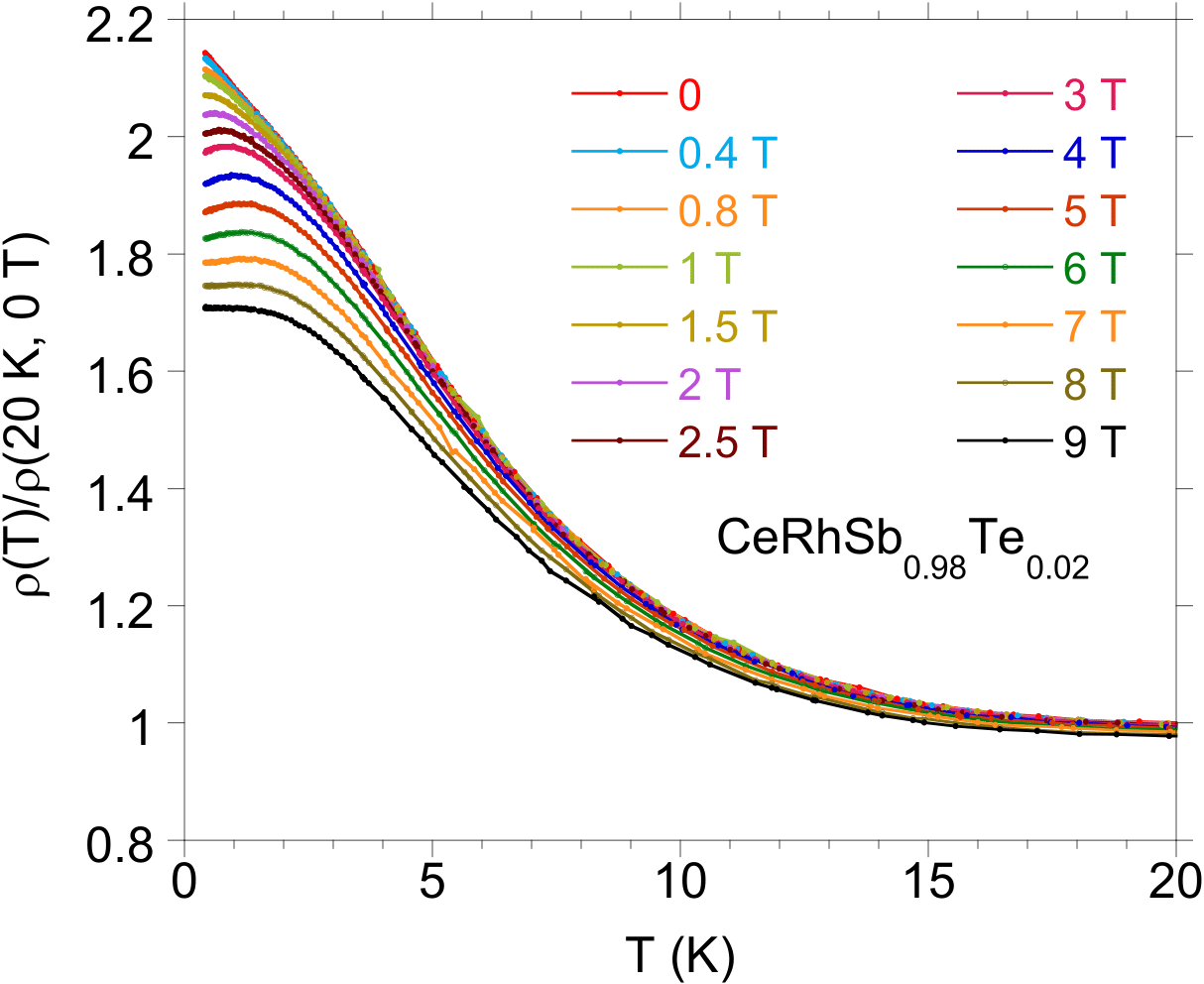}
\caption{\label{fig:Fig_R_Te-02}
CeRhSb$_{0.98}$Te$_{0.02}$. The low-temperature electrical resistivity $\rho(T)$, normalized to $\rho = 262$ $\mu\Omega$cm, at $T=20$ K and $B=0$, $\rho (T)/\rho (T=20 K)$, for different magnetic field values. 
} 
\end{figure}
Figure \ref{fig:Fig_R_Te-02} exhibits  low-temperature electrical resistivity $\rho$ at different magnetic fields for CeRhSb$_{0.98}$Te$_{0.02}$. The curves   show a well--defined thermally activated behavior, where $\rho \sim exp(\Delta/T)$ in the low-temperature regime that is characteristic of a nondegenerate  semiconductor with activated energy $\Delta(B)$, shown in Fig. \ref{fig:Fig_Gap-vs-B_x}. 
We should  also note that: (i)  an increase in $\rho$  generated by a KI (pseudo)gap  decreases systematically with increasing  $x$, (ii) the magnetic field leads to  the saturation of $\rho$ at  the lowest temperatures $T<\sim 2$ K, and finally (iii) in the applied field up to 9 T the  CeRhSb gap is reduced  to its of 95\%  value of $\Delta(0)$, whereas the doping leads to stronger  field reducing of $\Delta$, as illustrated  in Fig. \ref{fig:Fig_Gap-vs-B_x}. However, the field influence is not $x$-dependent and is weaker than expected.

\begin{figure}[ht]
\includegraphics[width=0.48\textwidth]{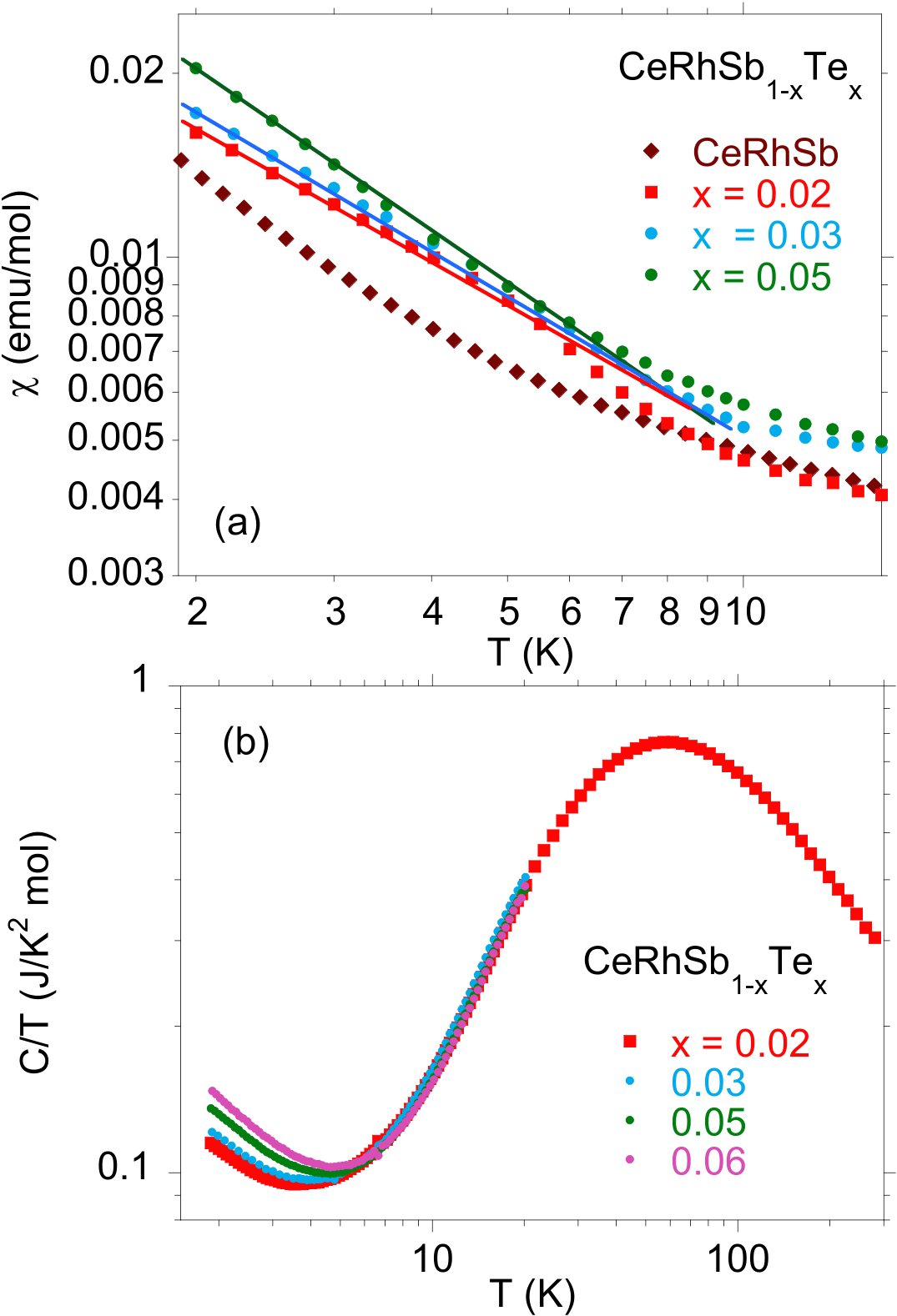}
\caption{\label{fig:Fig_CHI_C_Te_A-B}
CeRhSb$_{0.98}$Te$_{0.02}$: low temperature magnetic susceptibility (dc) vs $T$ in log-log scale (a) and the specific heat divided by temperature, $C/T$, in log-log scale. Both quantities for $x=0.02$, 0.03, 0.05, and 0.06 show a singular behavior at the low-$T$ region.
} 
\end{figure}

\begin{figure}[ht]
\includegraphics[width=0.48\textwidth]{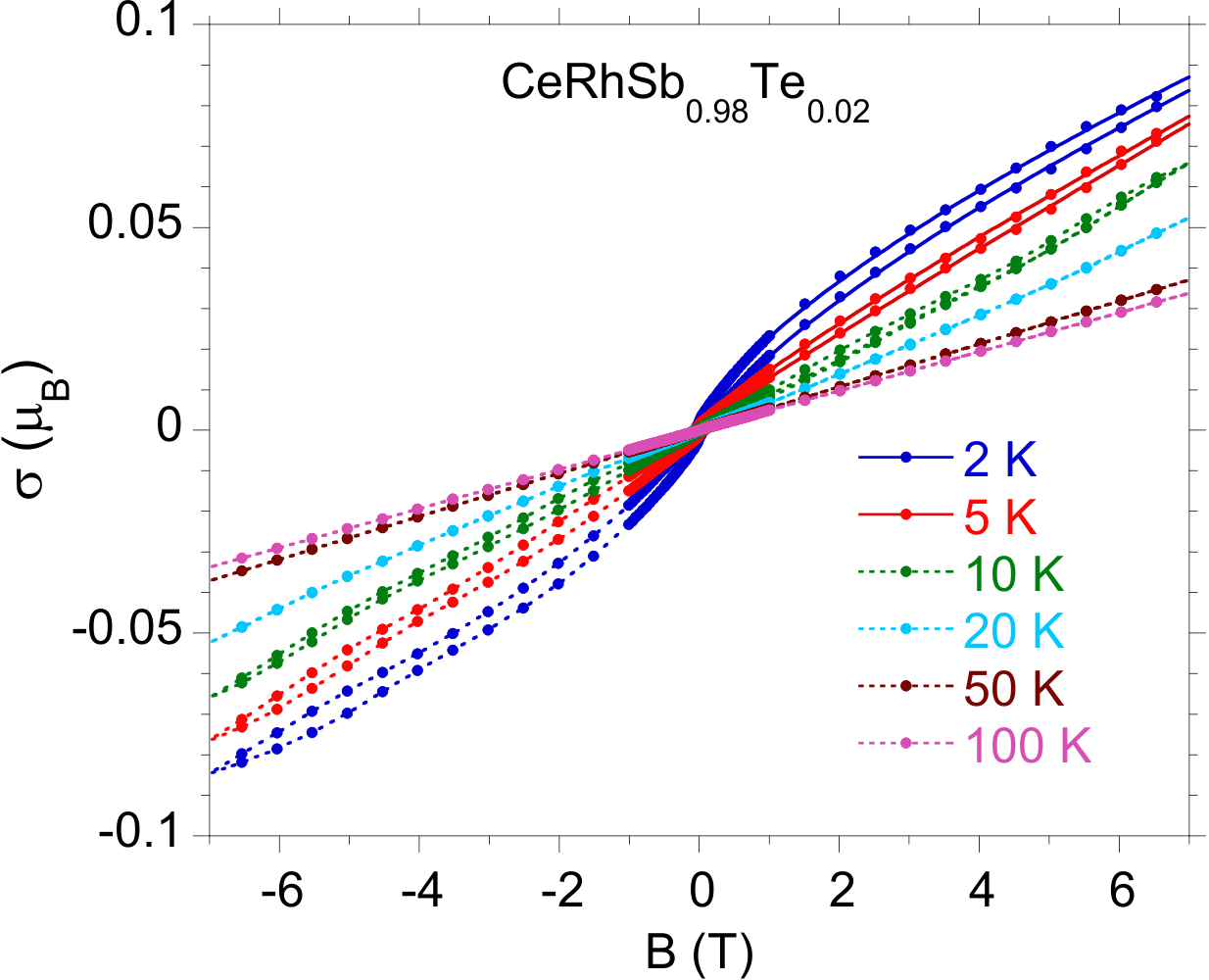}
\caption{\label{fig:Fig_M-H_Te_02_fit}
Magnetization $\sigma$ of CeRhSb$_{0.98}$Te$_{0.02}$  as a function of magnetic field, showing a hysteretic loop at 2 K and 5 K, and power law behavior, $\sigma \sim B^{\lambda}$. For each temperature the $\sigma$ isotherms were measured under  increasing and decreasing fields between -9 T and +9 T. 
The best fit with $\lambda=0.69$ is represented by the blue lines at $T=2$ K. For 5 K isotherm the best fit (red lines) gives $\lambda=0.95$. At the remaining temperatures ($T>10$ K) the hysteretic loop was not observed (the dotted lines connect the experimental points of the $\sigma$ isotherms).
} 
\end{figure}

\begin{figure}[ht]
\includegraphics[width=0.48\textwidth]{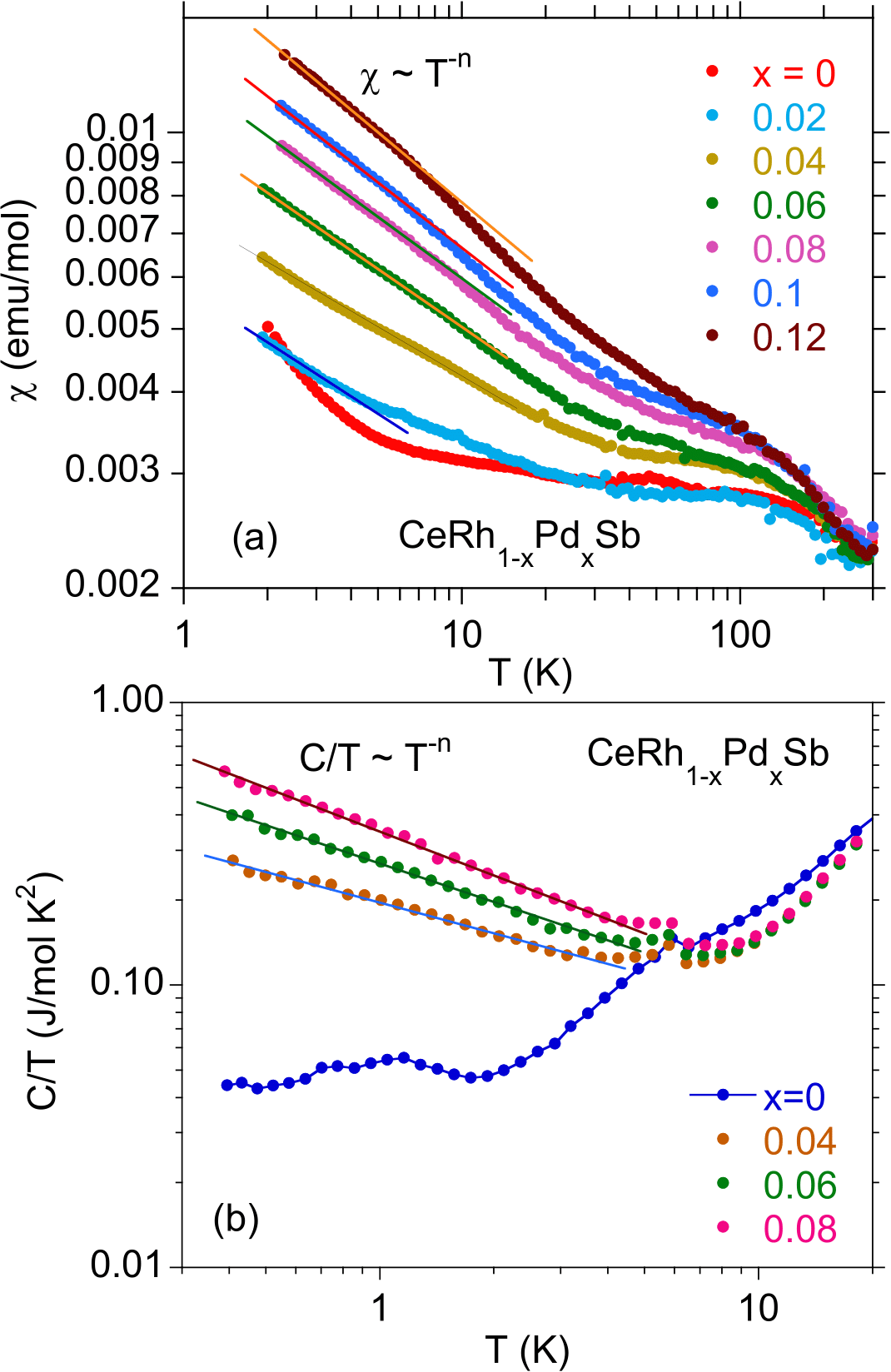}
\caption{\label{fig:Fig_CHI_C-Pd-Rh_A-B}
CeRh$_{1-x}$Pd$_x$Sb; low temperature magnetic susceptibility (dc) vs $T$ in log-log scale (a) and the specific heat divided by temperature, $C/T$, in log-log scale, with divergent behavior at the low-$T$ region (b).
A weak feature at 1.1 K  in $C/T$ of  CeRhSb indicates on the weak magnetic inhomogeneity, due to strongly diluted magnetic entities (see text).} 
\end{figure}

\begin{figure}[ht]
\includegraphics[width=0.48\textwidth]{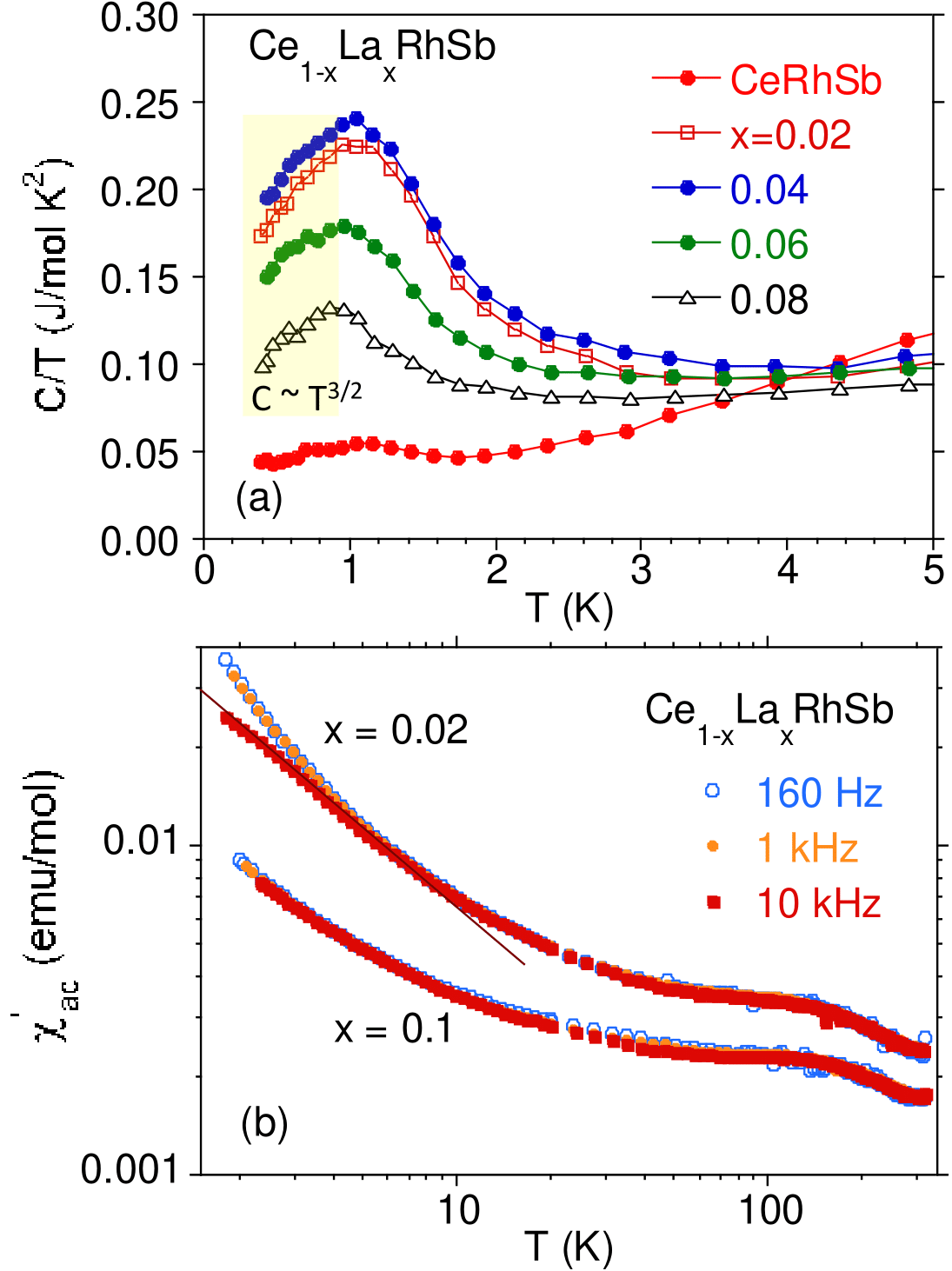}
\caption{\label{fig:Fig_C_CHI_f_A-B}
Temperature dependence of the specific heat divided by temperature, $C/T$, for Ce$_{1-x}$La$_x$RhSb (a). Panel (b) shows the ac susceptibility $\chi'$ vs $T$, at various frequencies.} 
\end{figure}

Low temperature magnetic susceptibility and $C/T$ data for CeRhSb$_{1-x}$Te$_{x}$ compounds ($x=0.02$, 0.03, and 0.05) are plotted on a double logarithmic scale in  Fig. \ref{fig:Fig_CHI_C_Te_A-B}(a). Both $\chi\sim T^{-1+\lambda}$ and $C/T \sim T^{-1+\lambda}$(b) follow a power-law temperature dependence with similar $\lambda$, as shown in Table \ref{tab:C_CHI}. 
The observed power law dependencies $\chi\sim T^{-1+\lambda}$ and $C/T\sim T^{-1+\lambda}$, and the required agreement between $\lambda$s from $C(T)/T$ and $\chi(T)$ data support the Griffiths phase model \cite{Castro1998}.  Moreover, magnetization as a function of field up to 7 T does follow the predicted Griffiths phase behavior $\sigma \sim B^{\lambda}$ \cite{Castro2000}, as shown in Fig. \ref{fig:Fig_M-H_Te_02_fit} and in Table \ref{tab:C_CHI}. Note that $\lambda$ from the best fit of expression $\sigma=\sigma(0)+mB^{\lambda}$  to the isotherms $\sigma_T(B)$,  is temperature dependent and decreases with $T\rightarrow 0$; the correct value of $\lambda$ is approximated by the $T=0$ value.
Likewise, Griffiths phase behavior has been previously documented for Ce$_{0.98}$Th$_{0.02}$RhSn (\cite{Kim2003}, Table \ref{tab:C_CHI}), and for the series of CeRh$_{1-x}$Pd$_x$Sb compounds (see Fig. \ref{fig:Fig_CHI_C-Pd-Rh_A-B}, Table \ref{tab:C_CHI}). Ce$_{0.98}$Th$_{0.02}$RhSn shows a non-Fermi-liquid behavior in the specific heat within two decades of temperature \cite{Kim2003}, which is consistent with a quantum critical point scenario. 
However, the low temperature specific heat divided by temperature, $C/T$, of Ce$_{0.98}$Th$_{0.02}$RhSn can also be fit  to a $T^{-1+\lambda}$  temperature dependence, which is consistent with the Griffiths phase disorder theory. Simultaneously, the low-$T$ magnetic susceptibility can also be expressed as a power law with  similar exponent $\lambda$ in a temperature range 0.15 K -- 10 K in agreement with the Griffiths phase model, and magnetization $\sigma \sim B^{\lambda}$.  
Note that the magnetization curves differ from a metamagnetic-like behavior of typical heavy fermion system \cite{Haen}. It is intriguing to observe that this is an additional feature, by which KIs can be differentiated from the standard metallic heavy fermions.  
Ce$_{1-x}$Th$_{x}$RhSn samples  with $x=$ 0.3 and 0.4  were  reported as  magnetic, 
with spin cluster freezing below $\sim 0.3$ K \cite{Kim2003}.  Similar magnetic Kondo-insulator instabilities  were  predicted theoretically  for the Kondo-hole systems, where they give  rise to a bound state in the gap, leading to a Stoner-like ferromagnetic instability (cf., \cite{Schlottmann1991a,Schlottmann1991b,Schlottmann1996}).
The Th and La impurities in CeRhSb are also regarded as Kondo holes. Clusters of Kondo holes have been studied in Ref. \cite{Slebarski2010}. Evidence of the spin-glass-like behavior is shown in Fig. \ref{fig:Chi_ac_f_CeLa02RhSb} for CeRhSb doped with La. 
Figure \ref{fig:Fig_C_CHI_f_A-B}(a) shows the spin-glass-like contribution  to the low-temperature specific heat $C/T$, obtained for the series of Ce$_{1-x}$La$_x$RhSb samples at $B=0$. Panel (b) shows the ac magnetic susceptibility $\chi'(T)$ measured at various frequencies $\nu$.  For La   impurities $\chi'$ basically  does not follow the power-law behavior, with exclusion of  the samples with  small La content, $x=0.02$ and 0.04, where the high frequency ($\nu=10$ kHz) real component of $\chi_{ac}$  has a
$\chi'\sim T^{-0.9}$ scaling between 1.9 K and 10 K [see Fig. \ref{fig:Fig_C_CHI_f_A-B}(b),  straight-line part]. The above described properties  suggest high-frequency-induced fluctuations of the  magnetic clusters which are blocked up below $T_f \sim 1$ K, as illustrated in Fig. \ref{fig:Chi_ac_f_CeLa02RhSb}. However, this is not the Griffits-phase case, as shown in Fig. \ref{fig:Fig_C_CHI_f_A-B}  $C\sim T^{3/2}$ for $T<T_f$, which could signal an appearance of an inhomogeneous magnetic state \cite{Coey1977}.
\begin{table*}[]
\caption{Specific heat $C/T$, d.c. magnetic susceptibility $\chi$ at 0.1 T, and magnetization  $\sigma$ parametrization of the results for Ce$_{1-x}$La$_x$RhSb, Ce$_{1-x}$Th$_x$RhSb \cite{Kim2003}, CeRh$_{1-x}$Ni$_x$Sb \cite{Menon1997}, CeRh$_{1-x}$Pd$_x$Sb and CeRhSb$_{1-x}$Te$_x$ samples within KI regime ($\Delta >0$). Susceptibility and specific heat, $C/T$, is fitted to a Griffiths phase scenario $\chi\sim C/T\sim T^{n}$, where $n=-1+\lambda$, and $\sigma \sim B^{\lambda}$ [$\lambda$ from $\sigma(B)$  isotherms is $T$ dependent and decreases when $T\rightarrow 0$].
$^{(a)}$ A small La doping $x\geqslant 0.02$ of CeRhSb leads to formation of spin-glass-like state \cite{Slebarski2010}.
$^{(b)}$ CeRh$_{1-x}$Pd$_x$Sb, the $\chi$, $C/T$, and $\sigma$ data indicate a presence of very weak magnetic correlations for the components $x=0.10$ and 0.12.
$^{(c)}$ Universal scaling law $\chi\rho=const.$ is documented  for the components $x<0.13$ of the series, after subtracting the  impurity contribution  $y\frac{C}{T}$ to  $\chi(T)$, $y< 0.01$ (see Ref. \cite{Spalek2005}).
}
\label{tab:C_CHI}
\begin{tabular}{ccccc}
\hline
compound  & $x$ & $\chi\sim T^{n}$ &  $C/T\sim T^{n}$ & $\sigma \sim B^{\lambda}$ \\
& &$n=-1+\lambda$ &$n=-1+\lambda$&   $\lambda$\\
\hline
Ce$_{1-x}$La$_x$RhSb &      &   &   &  \\                     
                     & $0< x\leq 0.14^{(a)}$  &not observed & not observed  &  \\
Ce$_{1-x}$Th$_x$RhSb \cite{Kim2003}&      &   &   &  \\ 
                                   & 0.02     & -0.36 & -0.26  & 0.69 ($T=2$ K) \\ 
                                   & 0.03     & -0.39 &   &  \\      
                                   & 0.04     & -0.44 &   &   \\
                                   &  & $T<10$ K & $0.05<T<1.3$ K &   \\
CeRh$_{1-x}$Ni$_x$Sb \cite{Menon1997}&   &   &   &  \\ 
                                   & 0.05 & -0.23 &   & \\
                                   & 0.1 & -0.22 &   & \\
                                   & $T<15$ K &  &   & \\ 
CeRh$_{1-x}$Pd$_x$Sb               &          &    &   &  \\  
                                   & 0.02 & -0.27 &  -0.20 & 0.83 ($T=2$ K) \\ 
                                   & 0.04 & -0.27 & -0.33 & 0.80 ($T=2$ K)  \\ 
                                   & 0.08 & -0.32 & -0.45  & 0.72 ($T=2$ K) \\ 
                                   & 0.10$^{(b)}$ & -0.38 &  -0.60 &  0.75 ($T=2$ K) \\  
                                   & 0.12$^{(b)}$ & -0.36 & -0.59  & 0.70 ($T=2$ K) \\ 
                                   &  & $T<10$ K &  $0.4<T<4$ K &  \\
CeRhSb$_{1-x}$Te$_x$               &      &    &   &  \\
                                   & 0.02 & -0.61 & -0.45 & 0.69 ($T=2$ K)\\  
                                   &      &      &      & $\sim 0.44$ ($T\rightarrow 0$)\\
                                   & 0.03 & -0.65 & -0.44 & 0.75 ($T=2$ K)\\ 
                                   &      &      &      & $\sim 0.60$ ($T\rightarrow 0$)\\  
                                   & 0.04 & -0.62 & -0.52 &  0.70 ($T=2$ K)      \\
                                  &      &      &      & $\sim 0.55$ ($T\rightarrow 0$)\\  
                                   & 0.05 & -0.70 & -0.50 & 0.67 ($T=2$ K)\\
                                   &      &      &      & $\sim 0.52$~($T\rightarrow 0$)\\
                                   & 0.06 &  -0.56    & -0.53 & 0.53 ($T=2$ K) \\            &      &      &      & $\sim 0.45$~($T\rightarrow 0$)\\      
                                   &  & $T<5$ K & $T<3.2$ K &   \\ 
CeRhSb$_{1-x}$Sn$_x$      &      &    &   &  \\                                                              &  $x\leq 0.12$    &  $\chi\rho =$ const$^{(c)}$  &  &  \\
                          &  0.13   &   -0.21 &  -0.33  & 0.71 ($T=2$ K) \\
                          &  0.14   &   -0.28 &       & 0.80 ($T=2$ K) \\
                          &  0.16   &   -0.10 &        & 0.78 ($T=2$ K) \\
                          &  0.19   &   -0.12 &        &  \\
                          &                  & $T<18$ K &  $T<4$ K &   \\ \hline
\end{tabular} 
\end{table*}

It is especially interesting when Sb in CeRhSb  is partially  substituted for Sn. This is the case with the decreasing  number of valence electrons in the system  CeRhSb$_{1-x}$Sn$_x$, when $x$ increases. Previously \cite{Spalek2005}, on the Sb-rich side ($0\leq x\leq 0.2$, with $x=0.2$ being the limit of solubility), we observe the transition KI$\rightarrow$NFL as a function of decreasing number of carriers. Namely, the system CeRhSb$_{1-x}$Sn$_x$ undergoes the Kondo insulator-weakly magnetic metal transformation via the quantum critical point (QCP) located at the critical concentration $x_{cr}\approx 0.12$. We have also observed  an accompanying universal scaling $\chi \rho = const.$ in the quantum-coherence regime  for CeRhSb$_{1-x}$Sn$_x$ with $x\leq 0.12$. In Table \ref{tab:C_CHI} we list the $\chi$, $\sigma$ and $C$ parametrizations of CeRhSb$_{1-x}$Sn$_x$. For $x>x_{cr}$, these quantities fit the Griffits-phase scenario for a non-Fermi-liquid. 

\subsection{Magnetoresistance of CeRhSb and \ch{CeRhSb_{0.98}Te_{0.02}} Kondo insulators}

\begin{figure}[ht]
\includegraphics[width=0.48\textwidth]{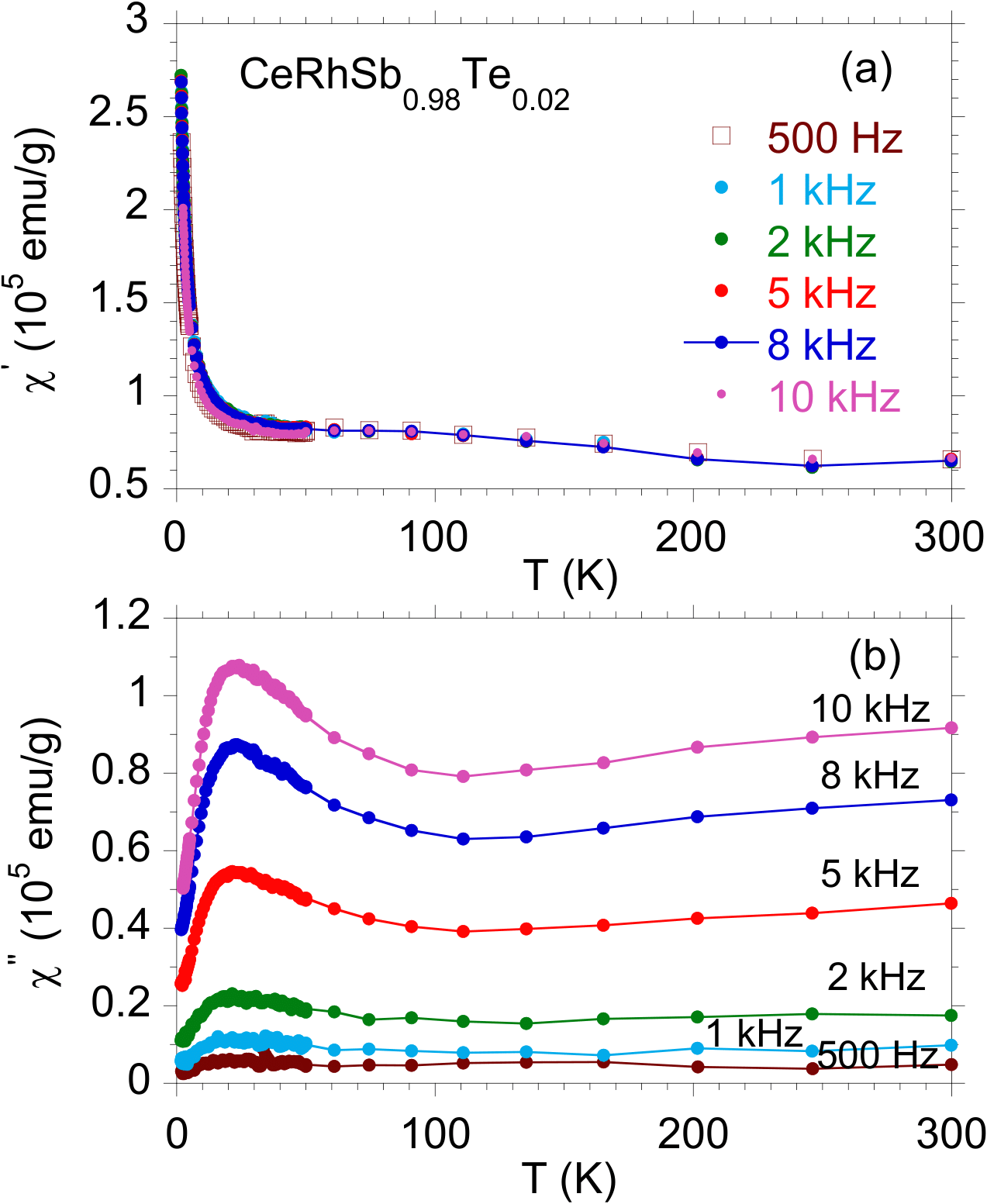}
\caption{\label{fig:chi-Te_2}
Temperature dependencies of the real $\chi'$ (a) and imaginary $\chi''$(b)  components of the ac mass magnetic susceptibility under magnetic field with amplitude of 2 G.  }
\end{figure}

Magnetoresistance, $MR$, of CeRhSb has been  reported previously. However, the various measurements for  obtaining  the $MR$ 
isotherms were limited  only to low temperatures (1.3 K \cite{Yoshino1998}, 1.7 K \cite{Adroja1994}), and to the magnetic 
field up to 4.5 T at 4.2 K at various pressure \cite{Uwatoko1994}, or were performed at 2 K, 4.5 K, and above 10 K under the 
increasing field with $B$ increment every $\sim 0.4$ T up to 5.5 T \cite{Malik1997}. Therefore, some interesting behaviors 
have been missed, especially in the $T$ region of the KI gap, $\Delta$.
\begin{figure}[ht]
\includegraphics[width=0.48\textwidth]{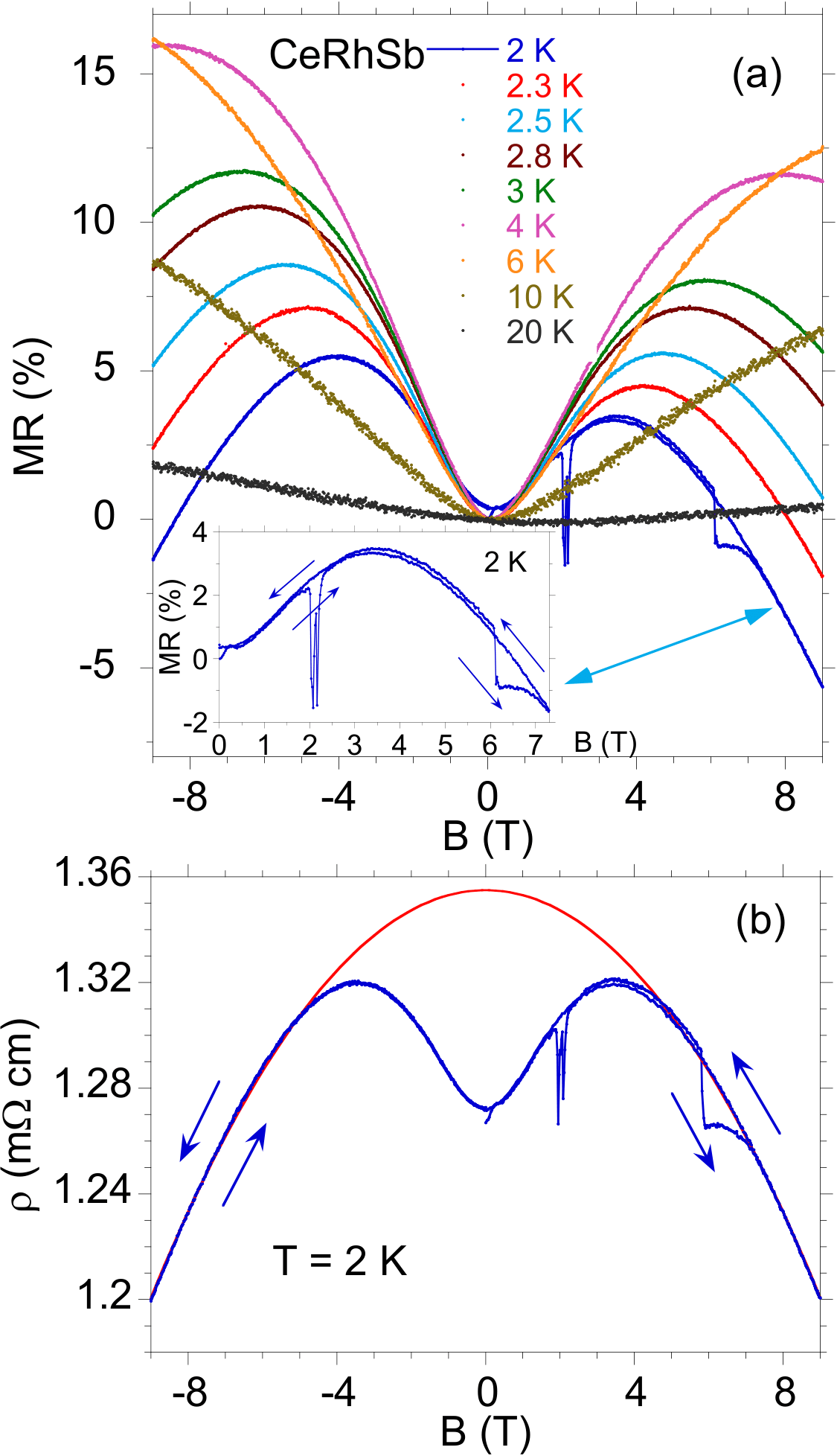}
\caption{\label{fig:MR_CeRhSb}
Magnetoresistance $MR= [\rho (B) - \rho (B = 0]/\rho (B = 0)\times 100$\%, of CeRhSb (sample thickness was 0.5 mm) as a function of the magnetic field, $B$, measured at several temperatures (a). The
arrows indicate field sweep directions. Magnetic field
sequence was ($0\rightarrow 9\rightarrow -9\rightarrow 0$) T. Inset displays details for 2 K $MR$ isotherm in the positive magnetic fields between 0 and 7 T. 
Panel (b) shows resistivity $\rho$ vs $B$   of CeRhSn at 2 K measured for the same sample, however  twice as thin (sample thickness was 0.5 mm). The field dependence of $\rho$ with  negative magnetoresistance effect presented  in panel (a) can be well fitted with the expression $\rho=\rho_0 + mB^2$, $m<0$ (red line).  }
\end{figure}
\begin{figure}[ht]
\includegraphics[width=0.48\textwidth]{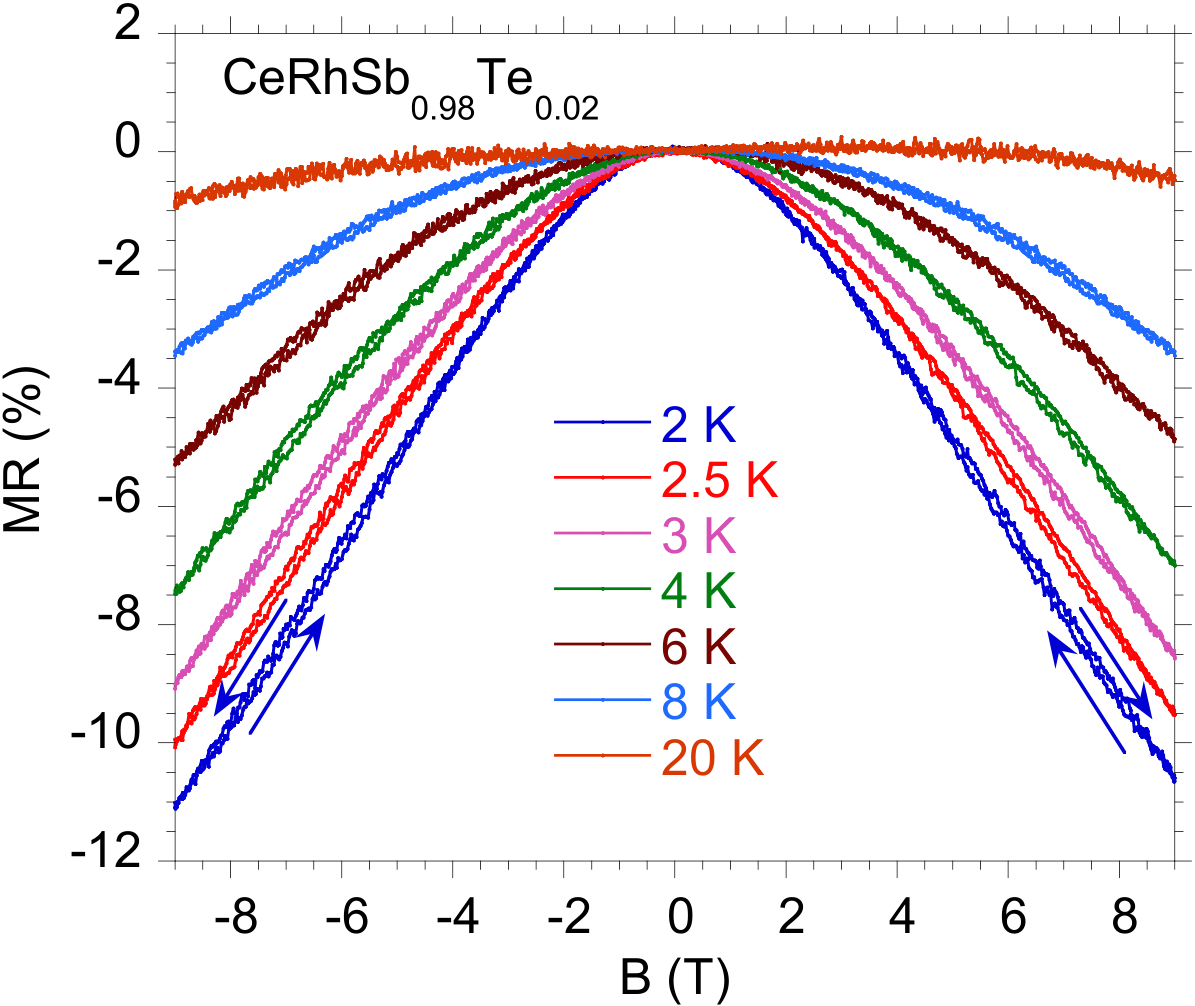}
\caption{\label{fig:Fig_MR_Te-02}
Magnetoresistance $MR = [\rho(B)- \rho(B = 0)]/\rho(B = 0)\times 100$\%, of CeRhSb$_{0.98}$Te$_{0.02}$ as a function of $B$, measured at several temperatures. The
arrows indicate field sweep directions. Magnetic field
sequence was ($0\rightarrow 9\rightarrow -9\rightarrow 0$) T (sample thickness was 0.5 mm).
} 
\end{figure}
Here, we have measured carefully the $MR$ isotherms  of CeRhSb and CeRhSb$_{0.98}$Te$_{0.02}$ at temperatures below, as well as above $\Delta/k_B$ in the fields up to 9 T, with magnetic field sequence ($0\rightarrow 9\rightarrow -9\rightarrow 0$) T,  as shown in  Figs. \ref{fig:MR_CeRhSb} and \ref{fig:Fig_MR_Te-02}.  First, we summarize the magnetoresistivity of parent CeRhSb: 
(i) Its  $MR$ isotherms at positive fields are in nature similar to those, reported  earlier, namely, at 2 K magnetoresistance in the increasing/decreasing directions of field sweeps shows a broad peak in the fields $B<6$ T, before it crosses over to
negative values for $B>6$ T. Such a positive $MR$  has also been previously observed at  low temperature  in, e.g.,  CeCu$_2$Si$_2$ and CeA1$_3$ heavy fermions, and was  discussed as  effect comming from   the quantum coherent nature of these Kondo lattices. 
On the other hand, the sign change in the magnetoresistance  at around 6 T indicates a disruption of the coherence and is also  characteristic feature of the Kondo lattice in the presence of local atomic  disorder in the vicinity of Ce ions. The negative magnetoresistance of CeRhSb at the  fields higher than 6 T can be  associated with the emergence of the local  $4f$ magnetic moments.
Okawa \cite{Ohkawa1990} explained this field induced phenomena theoretically.  He calculated 
the change of the low-temperature $MR_T$ isotherms from their positive to negative value  as a result of the presence of local defects around the Ce ions,  since local Kondo temperatures  are very sensitive to the atomic surrounding of the  $4f$  ions.  Within this theoretical approach the distribution  of the Kondo temperature leads to positive magnetoresistance at low fields, but negative $MR$ at high fields, as shown in Fig. \ref{fig:MR_CeRhSb}. For  temperatures $T> 2$ K crossover from positive  to negative value of MR is  expected for CeRhSb at much substantially higher field than 6 T, as illustrated in Fig. \ref{fig:MR_CeRhSb}.
(ii) Two   interesting features of the $MR$ isotherms at 2 K appear  at $B= 2$ T and 6 T with the increasing field sweep direction, which  disappear in the run of    field change in the remaining sequence of the measurement cycle  ($9 \rightarrow  -9 \rightarrow 0 $) T. These anomalies  disappear at  temperature higher than 2 K, as displayed in Fig. \ref{fig:MR_CeRhSb}. Note, these  field induced  anomalies are reproducible in repeated cycles of measurements of the $MR_{2 K}$ isotherm after cooling the sample. Nevertheless, only in the runs with the increasing field.
In general, doped both CeRhSb and CeNiSn are  disordered systems with a random anisotropy and competing interparticle interactions. Then, magnetic nanoparticles can be formed. There are known examples of Ce-Kondo lattices, doped  with nonmagnetic transition elements \cite{Slebarski2009},  produce a specific local electronic environment around the Ce ions. In effect, the  exchange interactions between the nearest Ce ions depend on the random occupation in the vicinity of each of them and  even could lead to a spin-glass-like state  at low temperature [so-called nonmagnetic atom disorder (NMAD) spin glass state \cite{Gschneidner1990}]. In a very diluted system, i.e., where the number of defects is small  and the interparticle dipole-dipole interactions are negligibly small in comparison with anisotropy energy,  the low temperature anomalous phenomena, attributed to dynamics of weakly magnetic nanoparticles, can be explained by superparamagnetic framework of the N\'eel-Brown model \cite{Neel1949,Brown1963}. 
Within the model, the flipping rate  $\tau$ for the nanomagnetic particle is$\tau=\tau_0 exp(-KV/k_BT)$, where  $\tau_0 $ is a microscopic relaxation time in the order of  ($10^{-9}-10^{-13}$) s,  $K$ is the anisotropy constant, and $V$ is the volume of particle. When $KV \gg k_BT$, the magnetic moment is frozen, that is the spin flipping is blocked \cite{comment1}. The blocking behavior is displayed in  Fig. \ref{fig:MR_CeRhSb}
at the field  2.05 T, 2.17 T, and 6.12 T, and corresponds  to nanoparticles with frequency $\nu=1/\tau_0=7.3\times 10^{10}$ s$^{-1}$,  $7.7\times 10^{10}$ s$^{-1}$, and $21.8\times 10^{10}$ s$^{-1}$, respectively  (assuming also that each magnetic cluster has  moment of 1 $\mu_B$). A probable scenario assumes  blocking of  nanoparticles with larger volume first and then the smaller next with the increasing $B$, whereas under  larger fields,  the particles are oriented with $B$ providing a negative $MR$ behavior. 
(iii) Figure  \ref{fig:Fig_MR_Te-02} exhibits  $MR$ vs $B$ for CeRhSb$_{0.98}$Te$_{0.02}$ at selected temperatures.
The $MR_T$  isotherms have  negative values at all fields and exhibit a weak hysteresis loops  between $\sim 4$ T and $\sim 8$ T for $T<6$ K.  Hovever, in this situation,  the fluctuating  nanoparticle moments are not correlated with each other. The component  $\chi_{ac}'$  shown in Fig. \ref{fig:chi-Te_2} does no depend on $\nu$, while $\chi_{ac}''$
 shows a strong  frequency dependence, different  than  that, measured for Ce$_{0.98}$La$_{0.02}$RhSb, characteristic of spin-glass-like phase (cf. Fig. \ref{fig:Chi_ac_f_CeLa02RhSb}).
The scenario with  noninteracting weakly magnetic nanoparticles  in  CeRhSb$_{0.98}$Te$_{0.02}$  correlates well with the observation of non-Fermi-liquid  $\chi \sim C/T \sim T^{-1+\lambda}$ $T$-dependencies, related to the presence of Griffiths phase in these  materials. 
(iv) In Fig. \ref{fig:MR_CeRhSb} the various $MR_T$ isotherms measured for  1 mm thick sample under positive fields,  are asymmetric in comparison to respective  $MR_T$ curves obtained in opposite-sign  field. The effect is due to the distribution of  the electric field between the electrodes and depends on the thickness of the sample. This is because for the  sample of 0.5 mm  thick the magnetoresistance at 2 K is fully symmetric under positive and negative fields. Moreover, the magnetic field dependence of the negative magnetoresistance can be  fitted accurately by the expression $\rho(B) = \rho(0) + mB^2$, $m<0$. Figure \ref{fig:Fig_MR_Te-02} shows a fully symmetric field dependence of $MR_T$ isotherms for CeRhSb$_{0.98}$Te$_{0.02}$  sample of thickness of $\sim 0.5$ mm. A negative magnetoresistance magnitude is increasing with decreasing  temperature.
To interpret the $MR$ isotherms shown in Figs. \ref{fig:MR_CeRhSb} and \ref{fig:Fig_MR_Te-02}, we suggest the following scenario.
It has been previously predicted that CeRhSb and CeNiSn could be  novel Kondo systems possessing the so-called M\"obius-twisted surface states (MKI) \cite{Chang2017}.  According to high resolution photoemission spectroscopy measurements \cite{Kumigashira2001,Shimada2002}, the both systems can be considered to be semimetallic with an anisotropic hybridization pseudogap. Thus, CeRhSb and CeNiSn are sometime called {\it failed Kondo insulators}. The resistivity of CeRhSb exhibits a tendency to saturate at $T\rightarrow 0$, as illustrated in Fig. \ref{fig:Fig_CeRhSb_CHI-R_skalowanie}. A pronounced  saturation effect was also detected for CeNiSn \cite{Takabatake1992,Slebarski2008} (the saturation of $\rho$ within the gap range is evidently documented for the off-stoichiometric CeNiSn \cite{Slebarski2009a}). The ARPES experiment \cite{Seong2019} did not confirm  unequivocally M\"obius topological surface states, predicted  for MKI CeRhSb for (010) surface,  but it has not been ruled out completely.
A. Burkov  \cite{Burkov2015} presented a microscopic model of diffusive magnetotransport in Weyl metals and has thus clarified its relation to
the chiral anomaly. Namely, he showed
that a measurable consequence of chiral magnetic effect  (CME) is a positive magnetoconductivity, proportional to $B^2$ in the limit of  weak magnetic field. In consequence,  Weyl metals possess  quadratic negative magnetoresistance, which could  dominate over other contributions,  as  was documented experimentally for various topological materials  (see, e.g., Refs. \cite{Cui2021,Boer2019,Hirschberger2016}).  
In summary, we  suggest  that the magnetoresistance properties  of CeRhSb analyzed here  may 
serve as  evidence of  the topologically nontrivial nature of this compound.
Furthermore, the negative magnetoresistance at 2 K with a bell-shaped
profile strongly
supports the emergence of the chiral anomaly in CeRhSb, as well as in CeRhSb$_{0.98}$Te$_{0.02}$, helping to rationalize the 
origin of the MR (cf., Figs.  \ref{fig:MR_CeRhSb} and \ref{fig:Fig_MR_Te-02}).

\section{Outlook}

In this paper, we have tested first the robustness of selected universal properties of Kondo insulators against the formation of defects and substitutional doping of systems based on \ch{CeRhSb} as the initial narrow--gap nonmagnetic semiconductor. In this respect, we tested the proposed by us original scaling $\chi \rho = const.$ and have discussed its limitations. Second, we have observed the onset of inhomogeneous magnetic stare appearing in the doped system \ch{CeRhSb_{1-x}Te_{x}} connected with the increased carrier concentration. The magnetic properties of the last compound provide clear evidence for a breakdown of the collective spin--singlet states with the minute doping. This feature obscures the pure Kondo insulating behavior as evidenced through the observed magnetic and electric (magneto)resistance behavior as a function of either temperature or applied magnetic field. The collected data concerning the new temperature and applied magnetic field scaling are summarized in Table \ref{tab:C_CHI}. On the basis of those scalings suggestions concerning the topological character of their states are made. 

\section{acknowledgments}

J.S. acknowledges the support from Narodowe Centrum Nauki, Grants OPUS No.~UMO-2018/29/B/ST3/02646 and No.~UMO-2021/41/B/ST3/04070.

\appendix
\section{Anderson--Kondo model for deriving the magnetic susceptibility}

The temperature dependence of magnetic susceptibility has been derived starting from Andreson--Kondo lattice Hamiltonian in the strong correlation limit, which has the form 

\begin{align}
    &\mathcal{H} = \hat{\mathcal{P}} \left\{ \sum_{mn\sigma}t_{mn}\hat{c}^{\dagger}_{m\sigma}\hat{c}_{n\sigma}  + \sum_{im\sigma} \frac{2|V_{im}|^2}{U+\epsilon_f} \frac{\hat{v}_{if\bar{\sigma}}}{4}\right\}\hat{\mathcal{P}} \\ \nonumber
    &+ \hat{\mathcal{P}} \left\{\sum_{i\sigma}\epsilon_f\hat{v}_{if\sigma} + \sum_{im\sigma}\left(V_{im}\hat{f}^{\dagger}\hat{c}_{m\sigma}\right) + \mathrm{H.c.} \right\}\hat{\mathcal{P}} \\ \nonumber
    &+ \hat{\mathcal{P}} \left\{ \sum_{im}\frac{2|V_{im}|^2}{U+\epsilon_f}\left(\hat{\textbf{S}}_i\cdot\hat{\textbf{s}}_m - \frac{\hat{v}_{if}\hat{n}_m}{4}\right)\right\}\hat{\mathcal{P}}  \\ \nonumber
    &+ \hat{\mathcal{P}} \left\{ -\frac{1}{2}g_f\mu_B H \sum_{i\sigma} \sigma \hat{N}_{i\sigma} - \frac{1}{2}g_c \mu_B H \sum_{m\sigma} \sigma \hat{n}_{m\sigma}\right\}\hat{\mathcal{P}} \\ \nonumber
    &+ \hat{\mathcal{P}} \left\{ \sum_{ij} J_{ij}\left(\hat{\textbf{S}}_i\cdot\hat{\textbf{S}}_j - \frac{\hat{v}_{if}\hat{v}_{jf}}{4}\right)\right\}\hat{\mathcal{P}}. \nonumber
\end{align}

In this Hamiltonian, the projection operators $\hat{\mathcal{P}}$ exclude double $f$--orbital occupancies. The consecutive terms are: The first represents the effective hopping between conduction--electron ($c$) sites ($m,n$) with $t_{mn}$ being the bare hopping amplitude, $V_{im}$ is the hybridization amplitude between $c$ and $f$ electrons.
The second is composed of spin--flip assistant $c$--electron hopping ($\hat{\textbf{S}}_i$ is the $f$--electron spin operator) and the bare atomic $f$--level position $\epsilon_f$. The third term represents the projected hybridization part. The fourth and the sixth terms represent the Kondo ($\tilde V^2/(U+\epsilon_f)$) and $f$ exchange 
($\tilde V^4/(U+\epsilon_f)^3$) parts, respectively. The fifth term represents the Zeeman energy in the static applied field $H$ for $f$ and $c$ electrons, respectively. Finally, $U$ is the interatomic $f-f$ Coulomb coupling. In general, this type of Hamiltonian takes into account Kondo ($f-c$) and superexchange interactions in an explicit form in addition to the intraatomic correlation. Such a starting point matches well with the subsequent solution which accounts accurately only for intraatomic correlations.

This Hamiltonian is solved in the statistically consistent Gutzwiller--approximation (SGA), in which the effective single--particle Hamiltonian takes the form, 
\begin{align}
    &\mathcal{H}_{MF} = -\sum_{<mn>\sigma}(\nu(\hat{c}^{\dagger}_{m\sigma}\hat{c}_{n\sigma} - \zeta) + \mathrm{H.c.})  \\ \nonumber
    &- \lambda_{cm}\sum_{m\sigma}(\sigma(\hat{c}^{\dagger}_{m\sigma}\hat{c}_{m\sigma} - m_c)) \\ \nonumber
    &- \mu \sum_{i\sigma}(\hat{c}^{\dagger}_{i\sigma}\hat{c}_{i\sigma} + \hat{f}^{\dagger}_{i\sigma}\hat{f}_{i\sigma}) \\ \nonumber
    &- \sum_{i\sigma}(\tau_{\sigma}(\hat{f}^{\dagger}_{i\sigma}\hat{f}_{i\sigma} - m_f)) \\ \nonumber
    &-\lambda\sum_{i\sigma} (\hat{f}^{\dagger}_{i\sigma}\hat{f}_{i\sigma} + \hat{c}^{\dagger}_{i\sigma}\hat{c}_{i\sigma} -n_e ) + <\mathcal{H}>, \nonumber 
\end{align}
where the variational parameters $\nu, \lambda_{cm}, \lambda_f, \lambda$, and the chemical potential $\mu$, are all determined variationally so that the statistical consistency is obeyed. For details of this cumbersome analysis see. \cite{HowczakSpalek,PhD-Howczak}
In the approach, the average $<\mathcal{H}>$ is approximated in such a manner that the effective mean--field Hamiltonian $\mathcal{H}_{mf}$ of single--particle form, with a number of effective fields to be evaluated through a self--consistent procedure, involving simultaneous solution up to 12 integral equations in the most general case.


\begin{thebibliography}{99}

\bibitem{Aeppli1992}
G. Aeppli and Z. Fisk, Comments Condens. Matter Phys. {\bf 16}, 155 (1992).

\bibitem{Ekino1995}
T. Ekino, T. Takabatake, H. Tanaka, and H. Fujii, Phys. Rev. Lett. {\bf 75}, 4262 (1995).

\bibitem{Spalek2000}
J. Spa\l ek, Acta Phys. Pol. A {\bf 97}, 71 (2000).

\bibitem{Takabatake1998}
 T. Takabatake, Y. Nakazawa, M. Ishikawa, T. Sakakibara, K. Koga, and I. Oguro, J. Magn. Magn. Mater. {\bf 77-76}, 87 (1988).  

 \bibitem{Slebarski1998}
 A. \'{S}lebarski, A. Jezierski, A. Zygmunt, S. M\"ahl, and M. Neumann, Phys. Rev. B {\bf 58}, 13498 (1998).

\bibitem{Slebarski1996}
A. \'{S}lebarski, A. Jezierski, A. Zygmunt, S. M\"ahl, M. Neumann, and G. Borstel, Phys. Rev. B {\bf 54}, 13551 (1996).

\bibitem{Dzero2010}
M. Dzero, K. Sun, V. Galitski, and P. Coleman, Phys. Rev. Lett. {\bf 104}, 106408 (2010).
\bibitem{Dzero2012}
M. Dzero, K. Sun, P. Coleman, and V. Galitski, Phys. Rev. B {\bf 85}, 045130 (2012).

\bibitem{Neupane2013}
M. Neupane, N. Alidoust, S-Y. Xu, T. Kondo, Y. Ishida, D. J. Kim, Chang Liu, I. Belopolski, Y. J. Jo, T-R. Chang, H-T. Jeng, T. Durakiewicz, L. Balicas, H. Lin, A. Bansil, S. Shin, Z. Fisk, and  M. Z. Hasan, Nat. Commun. {\bf 4}, 3991 (2013). 

\bibitem{Xu2013}
N. Xu, X. Shi, P. K. Biswas, C. E. Matt, R. S. Dhaka, Y. Huang, N. C. Plumb, M. Radovi\'{c}, J. H. Dil, E. Pomjakushina, K. Conder, A. Amato, Z. Salman, D. McK. Paul, J. Mesot, H. Ding, and M. Shi, Phys. Rev. B {\bf 88}, 121102(R) (2013).


\bibitem{Jiang2013}
J. Jiang, S. Li, T. Zhang, Z. Sun, F. Chen, Z.R. Ye, M. Xu, Q.Q. Ge, S.Y. Tan, X.H. Niu, M. Xia, B.P. Xie, Y.F. Li, X.H. Chen, H.H. Wen, and  D.L. Feng, Nat. Commun. {\bf 4}, 4010 (2013). 

\bibitem{Wolgast2013}
S. Wolgast, \c C. Kurdak, K. Sun, J. W. Allen, Dae-Jeong Kim, and Z. Fisk, Phys. Rev. B {\bf 88}, 180405(R) (2013).

\bibitem{Kim2013}
D. J. Kim, S. Thomas, T. Grant, J. Botimer, Z. Fisk, and  Jing Xia, Sci. Rep. {\bf 3}, 3150 (2013). 

\bibitem{Kim2014}
D. J. Kim, J. Xia, and  Z. Fisk, Nat. Mater. {\bf 13}, 466 (2014). 

\bibitem{Chang2017}
P. -Y. Chang, O. Erten, and P. Coleman, Nat. Phys. {\bf 13}, 794 (2017).

\bibitem{Nam2019}
T.-S. Nam, Chang-Jong Kang, D.-C. Ryu, Junwon Kim, Heejung Kim, Kyoo Kim, and B. I. Min, Phys. Rev. B {\bf 99}, 125115 (2019).

\bibitem{Nakamoto1995}
G. Nakamoto, T. Takabatake, H. Fujii, A. Minami, K. Maezawa, I. Oguro, and A. A. Menovsky, J. Phys. Soc. Jpn. {\bf 64}, 4834 (1995).

\bibitem{Seong2019}
Seungho Seong, Kyoo Kim, Eunsook Lee, Chang-Jong Kang, Taesik Nam, B. I. Min, Takenobu Yoshino, T. Takabatake, J. D. Denlinger, and J.-S. Kang, Phys. Rev. B {\bf 100}, 035121 (2019).

\bibitem{Malik1991}
S. K. Malik and D. T. Adroja, Phys. Rev. B {\bf 43}, 6295 (1991).

\bibitem{Slebarski2006a}
A. \'{S}lebarski, J. Alloys Compd. {\bf 423}, 15 (2006).

\bibitem{Slebarski2002a}
A. \'{S}lebarski, M. B. Maple, E. J. Freeman, C. Sirvent,
M. Rad\l owska, A. Jezierski, E. Granado, Q. Huang
and J. W. Lynn, Phil. Mag.  {\bf 82}, 943 2002.

\bibitem{Toby2006}
B. H. Toby, Powder Diffr. {\bf 21},67 (2006).

\bibitem{Slebarski2010}
A. \'Slebarski, J. Spa\l ek, M. Fija\l kowski, J. Goraus, T. Cichorek, and \L{}. Bochenek, Phys. Rev. B {\bf 82}, 235106  (2010).

\bibitem{Slebarski2005}
A. \'Slebarski and J. Spa\l ek, Phys. Rev. Lett. {\bf 95},  046402  (2005).

\bibitem{Spalek2005}
J. Spa\l ek, A. \'Slebarski, J. Goraus, L. Spa\l ek, K. Tomala, A. Zarzycki, 
and A. Hackemer, Phys. Rev. B {\bf 72}, 155112 (2005); A. \'Slebarski, J. Spa\l ek, M. Gam\.za, and A. Hackemer, Phys. Rev. B {\bf 73}, 205115 (2006).

\bibitem{Slebarski2008}
A. \'Slebarski, M.B. Maple, R.E. Baumbach, and T.A. Sayles, Phys. Rev. B {\bf 77}, 245133 (2008).

\bibitem{Slebarski2009a}
A. \'Slebarski A and J. Spa\l ek,  Phil. Mag. {\bf 89}, 1845 (2009). 

\bibitem{Spalek2011}
J. Spa\l ek and A. \'Slebarski, J. Phys.: Conf. Ser. {\bf 273}, 012055 (2011).

\bibitem{Hundley1990}
M. F. Hundley, P. C. Canfield, J. D. Thompson, Z. Fisk, and J. M. Lawrence, Phys. Rev. B {\bf 42}, 6842 (1990). 

\bibitem{Jaccarino1967}
V. Jaccarino, G. K. Wertheim, J. H. Wernick, L. R. Walker,  and A. Sigurds, Phys.  Rev.  {\bf 160}, 476 (1967); G. Aeppli and Z. Fisk, Comments Condens. Matter Phys. {\bf 16}, 155 (1992). 

\bibitem{Takabatake1995}
T. Takabatake, T. Yoshino, H. Tanaka, Y. Bando, H. Fujii, T. Fujita, H. Shida, and T. Suzuki, Physica B {\bf 206--207}, 804 (1995).

\bibitem{PhD-Howczak}
O. Howczak, Ph.D. thesis, Jagiellonian University, Krak\'{o}w 2012,  \url{https://th-www.if.uj.edu.pl/ztms/download/phdTheses/Olga_Howczak_doktorat.png}

\bibitem{Malik1995}
S. K. Malik, L. Menon, K. Ghosh, and S. Ramakrishnan
Phys. Rev. B {\bf 51}, 399 (1995).

\bibitem{Adroja1996}
D. T. Adroja, B. D. Rainford, A. J. Neville, P. Mandel, and A. G. M. Jansen, J. Magn. Magm. Mater. {\bf 161}, 157 (1996).

\bibitem{Kim2003}
J. S. Kim, E.-W. Scheidt, D. Mixson, B. Andraka, and G. R. Stewart, Phys. Rev. B {\bf 67}, 184401 (2003).

\bibitem{Menon1998}
L. Menon, F. E. Kayzel, A. de Visser, and S. K. Malik,
Phys. Rev. B {\bf 58}, 85 (1998).

\bibitem{Slebarski2010a}
A. \'Slebarski and J. Goraus, Phys. Stat. Solidi B {\bf 247},710 (2010).


\bibitem{Doradzinski1997}
R. Doradzi\'{n}ski and J. Spa\l ek, Phys. Rev. B {\bf 56},  R14239  (1997); Phys. Rev. B {\bf 58},  3293  (1998); for brief review see: J. Spa\l ek and R. Doradzi\'nski, Acta Phys. Polon. A {\bf 96}, 677 (1999); {\it ibid}. {\bf 97}, 71 (2000).

\bibitem{Anderson1961}
P. W. Anderson, Phys. Rev. {\bf 124}, 41 (1961).

\bibitem{Fuggle1983}
O. Gunnarsson and K. Sch\"onhammer, Phys. Rev. B {\bf 28}, 4315 (1983);
J.C. Fuggle, F.U. Hillebrecht, Z. Zolnierek, R. L\"asser, Ch.
Freiburg, O. Gunnarsson, and K. Sch\"onhammer, Phys. Rev. B,  {\bf 27}, 7330 (1983).

\bibitem{Castro1998}
A. H. Castro Neto, G. Castilla, and B. A. Jones
Phys. Rev. Lett. {\bf 81}, 3531 (1998). 

\bibitem{Castro2000}
A. H. Castro Neto and B. A. Jones
Phys. Rev. B {\bf 62}, 14975 (2000).

\bibitem{Haen}
P. Hean, J. Flouquet, F. Lappierre, P. Lejay, and G. Remenyi, J. Low Temp. Phys. {\bf 67}, 391 (1987).


\bibitem{Schlottmann1991a}
R. Sollie and P. Schlottmann, J. Appl. Phys. {\bf 69}, 5478 (1991).

\bibitem{Schlottmann1991b}
R. Sollie and P. Schlottmann, J. Appl. Phys. {\bf 70}, 5803 (1991).

\bibitem{Schlottmann1996}
P. Schlottmann, Phys. Rev. B {\bf 54}, 12324 (1996).

\bibitem{Coey1977}
J. M. D. Coey, S. vMolnar, and R. J. Gambino, Solid State Commun. {\bf 24}, 167 (1977).

\bibitem{Menon1997}
L. Menon and S. K. Malik, Physica B {\bf 230-232}, 695 (1997).

\bibitem{Yoshino1998}
T. Yoshino, T. Takabatake, M. Sera,
M Hiroi, N Takamoto, and K Kindo, J. Phys. Soc. Jpn. {\bf 67}, 2610 (1998).

\bibitem{Adroja1994}
D. T. Adroja and B. D. Rainford, J. Magn. Magn. Matter. {\bf 135}, 333 (1994).

\bibitem{Uwatoko1994}
Y. Uwatoko, G. Oomi, S. K. Malik, T. Takabatake, and H. Fujii, Physica B {\bf 199--200}, 572 (1994).

\bibitem{Malik1997}
S. K. Malik, L. Menon, V. K. Pecharsky, and K. A. Gschneidner, Phys. Rev. B {\bf 55}, 11471 (1997).



\bibitem{Ohkawa1990}
F. J. Ohkawa, Phys. Rev. Lett. {\bf 64}, 2300 (1990).


\bibitem{Slebarski2009}
A. \'{S}lebarski, J. Alloys Compd. {\bf 480}, 9 (2009).


\bibitem{Gschneidner1990}
K. A. Gschneidner Jr., J. Tang, S. K. Dhar, and A. Goldman, Physica B {\bf 163}, 507 (1990).

\bibitem{Neel1949}
L. N\'eel, Ann. Geophys. (C.N.R.S.) {\bf 5}, 99 (1949).

\bibitem{Brown1963}
W. F. Brown,Jr., Phys. Rev. {\bf 130}, 1677 (1963).

\bibitem{comment1}
At high temperatures , the moment flipping between two directions of its easy axis is described by superparamagnetism.

\bibitem{Kumigashira2001}
H. Kumigashira, T. Takahashi, S. Yoshii, and M. Kasaya
Phys. Rev. Lett. {\bf 87}, 067206 (2001). 

\bibitem{Shimada2002}
K. Shimada, K. Kobayashi, T. Narimura, P. Baltzer, H. Namatame, M. Taniguchi, T. Suemitsu, T. Sasakawa, and T. Takabatake, Phys. Rev. B {\bf 66}, 155202 (2002).


\bibitem{Takabatake1992}
T. Takabatake, M. Nagasawa, H. Fujii, G. Kido, M. Nohara, S. Nishigori, T. Suzuki, T. Fujita, R. Helfrich, U. Ahlheim, K. Fraas, C. Geibel, and F. Steglich, Phys. Rev. B {\bf 45}, 5740 (1992).


\bibitem{Burkov2015}
A. A. Burkov, Phys. Rev. B {\bf 91}, 245157 (2015).

\bibitem{Cui2021}
Y. Cui, Y. Chu, Z. Pan, Y. Xing, S. Huang, and
H. Xu, Nanoscale, {\bf 13}, 20417 (2021).

\bibitem{Boer2019}
J. C. de Boer, D. P. Leusink, and A. Brinkman, J. Phys. Commun. {\bf 3}, 115021 (2019).

\bibitem{Hirschberger2016}
M. Hirschberger1, S. Kushwaha, Z. Wang, Q. Gibson, S. Liang,
C. A. Belvin1, B. A. Bernevig, R. J. Cava, and N. P. Ong, Nature Materials {\bf 15}, 1161 (2016).

\bibitem{HowczakSpalek}
O. Howczak and J. Spa\l{}ek J. Phys.: Condens. Matter {\bf 24}, 205602 (2012)








\end{thebibliography}

\newpage

\end{document}